\documentclass[aps,prl,preprint,superscriptaddress]{revtex4-1}
\usepackage{amsmath}
\usepackage[pdftex]{graphicx}
\usepackage[thinspace]{SIunits}

\newcommand{\ket}[1]{\left| #1\right\rangle}

\providecommand \BibitemShut  [1]{\csname bibitem#1\endcsname}%

\begin{document}


\title{All optical quantum control of a spin-quantum state and ultrafast transduction into an electric current}

\author{K. M\"uller}
 \affiliation{Walter Schottky Institut and Physik-Department, Technische Universit\"at M\"unchen, Am Coulombwall 4, 85748 Garching, Germany \\}
\author{T. Kaldewey}
 \affiliation{Walter Schottky Institut and Physik-Department, Technische Universit\"at M\"unchen, Am Coulombwall 4, 85748 Garching, Germany \\}
\author{R. Ripszam}
 \affiliation{Walter Schottky Institut and Physik-Department, Technische Universit\"at M\"unchen, Am Coulombwall 4, 85748 Garching, Germany \\}
\author{J. S. Wildmann}
 \affiliation{Walter Schottky Institut and Physik-Department, Technische Universit\"at M\"unchen, Am Coulombwall 4, 85748 Garching, Germany \\}
 \author{A. Bechtold}
 \affiliation{Walter Schottky Institut and Physik-Department, Technische Universit\"at M\"unchen, Am Coulombwall 4, 85748 Garching, Germany \\}
\author{M. Bichler}
 \affiliation{Walter Schottky Institut and Physik-Department, Technische Universit\"at M\"unchen, Am Coulombwall 4, 85748 Garching, Germany \\}
 \author{G. Koblm\"uller}
 \affiliation{Walter Schottky Institut and Physik-Department, Technische Universit\"at M\"unchen, Am Coulombwall 4, 85748 Garching, Germany \\}
\author{G. Abstreiter}
 \affiliation{Walter Schottky Institut and Physik-Department, Technische Universit\"at M\"unchen, Am Coulombwall 4, 85748 Garching, Germany \\}
\author{J.J. Finley}
 \email{finley@wsi.tum.de}
 \affiliation{Walter Schottky Institut and Physik-Department, Technische Universit\"at M\"unchen, Am Coulombwall 4, 85748 Garching, Germany \\}

\date{\today}

%\keywords{}

\maketitle

%Introductory paragraph (150 words maximum)
\textbf{The ability to control and exploit quantum coherence and entanglement drives research across many fields ranging from ultra-cold quantum gases to spin systems in condensed matter. Transcending different physical systems, optical approaches have proven themselves to be particularly powerful, since they profit from the established toolbox of quantum optical techniques, are state-selective, contact-less and can be extremely fast. Here, we demonstrate how a precisely timed sequence of monochromatic ultrafast ($\sim2-5$ ps) optical pulses, with a well defined polarisation can be used to prepare arbitrary superpositions of exciton spin states in a semiconductor quantum dot, achieve ultrafast control of the spin-wavefunction \emph{without} an applied magnetic field and make high fidelity read-out the quantum state in an arbitrary basis simply by detecting a strong ($\sim2-10$ pA) electric current flowing in an external circuit. The results obtained show that the combined quantum state preparation, control and read-out can be performed with a \emph{near-unity} ($\geq97\%$) fidelity. Our methods are fully applicable to other quantum systems and have strong potential for scaling to more complex systems such as molecules and spin-chains.} 

Since the first proposals to use localised spins in solids as prototype quantum systems \cite{Loss1998, Imamoglu1999} much progress has been made using both electrons and holes in semiconductor quantum dots (QDs). Methods such as optical pumping \cite{Mete2006, Kim2010, Greilich2011, Weiss2012} and selective tunnel ionisation of spin polarised carriers \cite{Kroutvar2004,Heiss2007,Heiss2008,Godden2010, Godden2012, Ramsay08, Mueller2012} have been developed for reliable quantum state preparation and sensitive state readout can now be performed via spin-selective resonant fluorescence \cite{Vamivakas2009, Vamivakas2010} or absorption \cite{Heiss2008, Jovanov2011, Ramsay08, Mueller2012}. Recently, magnetic fields have been applied to achieve all optical coherent control of spin quantum states, for both large ensembles of dots\cite{Greilich2009} and single spins \cite{Press2008, DeGreve2011, Godden2012}. Whilst electron and hole spins have long $T_1$ lifetimes \cite{Kroutvar2004, Heiss2007} and robust quantum coherence ($T_2^{(*)}$) extending into the microsecond range \cite{Godden2012, Greilich2011, Xu2008} the neutral exciton spin has the advantage that arbitrarily polarised optical pulses can be used for quantum state control \cite{Kodriano2012}. However, exciton spin coherence persists over comparatively short timescales that are fundamentally limited by spontaneous emission ($\leq 1$ ns) \cite{Borri01}, necessitating ultrafast coherent control and readout. 

Very recently, coherent control of the exciton spin has been demonstrated at zero magnetic field by utilising the e-h exchange coupling to provide exciton spin rotations upon applying $2\pi$-optical pulses to excited biexciton transitions in single dots \cite{Poem2011, Kodriano2012} or by utilising e-e exchange interactions in QD-molecules \cite{ Kim2008}. All experiments reported to date require several laser control fields that are finely tuned to different transitions of the charged exciton manifold, calling for careful characterisation of each quantum sub-system and / or precise control of interdot couplings. 

Here, we describe how a sequence of three \emph{fully resonant} optical pulses, each having a precisely controlled optical polarisation, can be used to (i) prepare an arbitrary exciton spin superposition in an individual semiconductor QD \emph{without} a magnetic field, (ii) perform high-fidelity arbitrary spin rotations on the Bloch sphere and (iii) read out the spin wavefunction in an arbitrary measurement basis. The weak electron-hole exchange coupling ($\Omega_{eh}=10-40\mu$eV) that exists for neutral excitons in III-V quantum dots \cite{Finley2002,Xu2008} is shown to facilitate ultrafast spin control without the need to apply external magnetic fields by accumulating a polarisation dependent geometric phase during the presence of the $2\pi-$control pulse. Futhermore, we precisely measure the quantum spin state with unprecedented fidelity simply by detecting a large ($\sim3-5$ pA) electric current flowing in an external circuit attached to the dot. All-optical, ultrafast, high-fidelity spin preparation, control and read-out is demonstrated in an arbitrary basis with near unity fidelities ($\geq97\%$) limited only by the $\sim80$ fA electrical noise in the readout circuit. 

The sample investigated consists of a layer of low density self-assembled InGaAs quantum dots embedded into the intrinsic region of a electrically tunable n-i-Schottky diode \cite{Findeis2001,Zrenner02} (see supplementary information). This device geometry facilitates complementary photoluminescence (PL) emission and photocurrent (PC) absorption measurements by varying the applied electric field \cite{Jovanov2011,Fry2000,Findeis2001}. As described in the supplementary information, we derive three independent ps-pulse trains from a single $\sim100$ fs pulse that are individually tunable in energy, pulse area and relative time delay.  The polarisation state of each pulse is precisely controlled using liquid crystal variable retarders and waveplates.  By tuning the electric field very close to the transition from the PL to the PC regime ($\geq20$ kV/cm \cite{Mueller2011}) we detect the photocurrent $I$ induced in the sample by allowing carriers to tunnel out of the dot over timescales of several hundred picoseconds \cite{Mueller2012-1, Mueller2012} using a low-noise current-voltage converter. Blocking and unblocking the pump beam reveals the pump-induced change of the probe induced photocurrent $\Delta I$. The quantities $I$ and $\Delta I$ can be interpreted as the absorption of the quantum dot and its pump-induced change, respectively.

\begin{figure}
\includegraphics[width=1\columnwidth]{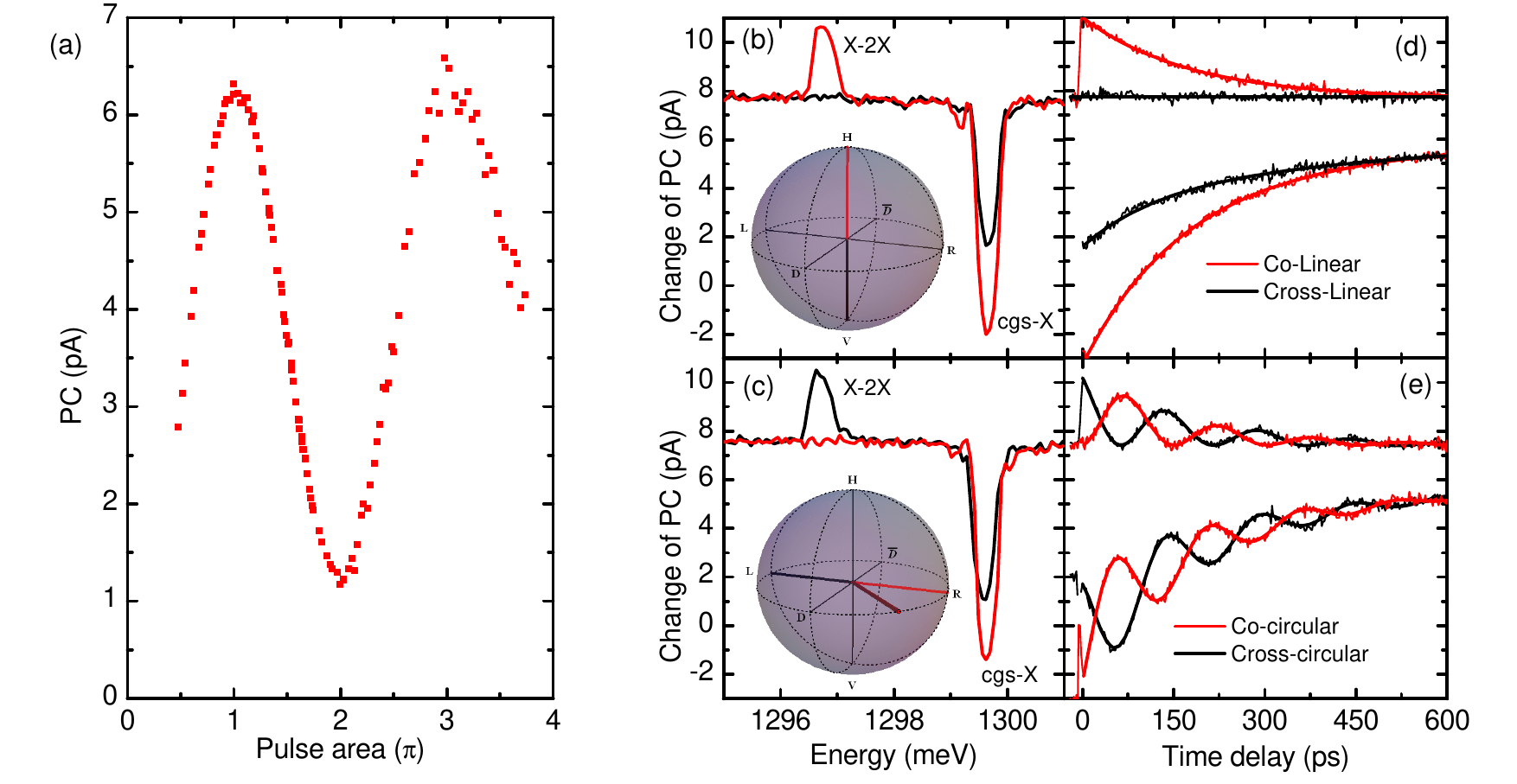}
\caption{\label{fig:Figure1}
(Color online)(a)Coherent photoresponse of the system upon driving the $cgs-X$ transition. (b-e) Pump-probe spectra recorded by pumping the $cgs-X$ transition for co- (cross-) linear configurations along the  $[110]$ and $[1\bar{1}0]$ crystal axes.. (d)Temporal evolution of the peaks/dips from (b). (c)and (e) Same experiment reported in (b) and (d) but for  for co- (cross-) \textit{circular} polarisations}
\end{figure}

We begin by identifying the crystal ground state ($cgs$) to neutral exciton ($X$) transition $cgs-X$ in both PL and single pulse PC measurements. All data presented in this manuscript were recorded at 4.2K for a fixed applied electric field of $27.8 \, kV/cm$.  Under these conditions, the $cgs-X$ transition is at $1299.6 \, meV$ for the dot investigated. Figure 1a shows the amplitude of the photocurrent induced by a single, $\sim6$ ps duration pump pulse tuned to $cgs-X$ as a function of the pulse area. Very clean Rabi oscillations are observed \cite{Zrenner02,Ramsay08} calibrating the laser power needed to fully invert the $cgs-X$ transition ($\pi$-pulse) and clearly demonstrating the coherent nature of the interaction between the driving laser field and the electrically contacted QD studied. We then continued to perform \emph{two pulse} experiments where the energy of the first (pump) pulse is tuned to $cgs-X$ and the pulse area is carefully set to $\pi$.  The second (probe) pulse is then tuned over a $\sim6$meV window in the immediate spectral vicinity of $cgs-X$. The result of such a pump-probe experiment is presented in fig. 1b that shows $\Delta I$ as a function of the probe pulse energy for a time delay of $5 \, ps$ between pump and probe. In this experiment, the polarisation of the pump pulse was set to be linear, aligned along the $[110]$ crystal axis and the probe pulse was set to be either co-linear or cross-linear to the pump for the red and black curves presented in the figure, respectively. The data clearly shows a pump induced bleaching\cite{Mueller2012} of the $cgs-X$ transition at $1299.6 \, meV$ (negative going dip) and conditional absorption (positive going peak at $1296.7 \, meV$) arising from the $X-2X$ transition, red detuned by $-2.9 \, meV$ \cite{Zecherle2010, Ramsay08, Mueller2012-1} from $cgs-X$. The conditional absorption of the $X-2X$ transition is only present for co-linear polarisations (red) and is entirely absent for the cross-linear configuration. Careful examination of fig. 1b shows that the bleaching of the $cgs-X$ transition is much stronger for the co-linear configuration, as compared with the cross linear polarisation. These findings can be explained as follows: The polarisation state of the exciton can be represented on the Bloch sphere as illustrated schematically in the inset of fig. 1b,  whereby the upper and lower states are the horizontal ($H$) and vertical ($V$) energy eigenstates of the exciton defined by the anisotropic e-h exchange coupling in the exciton\cite{Finley2002}. All states on the Bloch sphere can be written as coherent superpositions of these states. In terms of the single particle spin eigenstates $\uparrow$ and $\downarrow$ for electrons ($\Uparrow$ and $\Downarrow$ for holes) $H$ and $V$ can be written as $1/\sqrt{2}(\ket{\downarrow\Uparrow}+\ket{\uparrow\Downarrow})$ and $-i/\sqrt{2}(\ket{\downarrow\Uparrow-\uparrow\Downarrow})$ and an arbitrary spin wave function is defined with the phase factors $\phi$ and $\theta$, with a wavefunction $\Psi=(\cos(\frac{\theta}{2})\ket{H}+e^{i\phi/2}\sin(\frac{\theta}{2})\ket{V})$ . The optically active spin basis states that are represented by Bloch vectors along the $\pm y$ and $\pm x$ axis of the Bloch sphere are $R=\ket{\downarrow\Uparrow}$, $L=\ket{\uparrow\Downarrow}$ and $D=e^{-i\pi/4}/\sqrt{2}(\ket{\downarrow\Uparrow}+i \ket{\uparrow\Downarrow})$, $\bar{D}=e^{i\pi/4}/\sqrt{2}(\ket{\downarrow\Uparrow}-i \ket{\uparrow\Downarrow})$, respectively \cite{Benny2011}. Due to the optical selection rules for interband optical transitions, pulses resonant with $cgs-X$ having $\sigma^+$ ($\sigma^-$) polarisaion create excitons in the spin states $R$ ($L$) and linear polarisations parallel (perpendicular) to the [110] crystal axis will directly excite the spin eigenstates $H$ ($V$) in the presence of anisotropic $e-h$ exchange interaction\cite{Finley2002}. The polarisation state of the optical pulses on the Poincar\'e sphere is directly \emph{mapped} onto the exciton spin. For the data presented in fig. 1b we created an exciton in the state $H$. Therefore, the biexciton can only be created using a co-linear polarisation since the conditional absorption for cross-linear polarisations is Pauli spin forbidden.  In comparison, the strong bleaching of the $cgs-X$ transition in fig. 1b arises for cross-linear polarisations due to the fact that, as long as the dot is occupied by an electron and/or hole, the probe pulse cannot be absorbed. For co-linear polarisations (fig. 1b - red curve) the probe pulse causes stimulated emission resulting in a further reduction of $\Delta I$ and a deeper dip. Figure 1c shows data recorded for the same experiment as fig. 1b, but repeated for co- (cross-) \emph{circular} polarisations of pump and probe. Since stimulated emission is only possible if the polarisation of the probe pulse matches the exciton spin, the bleaching in fig. 1c is much stronger for co-circular polarisations as expected.  Similarly, the biexciton can only be created using cross-circular polarisations as evidenced by the absence of the $X-2X$ peak for the co-circular polarisation configuration of pump and probe. Figures 1d and 1e show the evolution of $\Delta I$ with probe pulse tuned to the $X-2X$ and $cgs-X$ transitions, respectively. As the time delay between pump and probe increases, both the conditional absorption of $X -2X$ and bleaching of $cgs-X$ transitions decrease exponentially due to tunnelling of electrons and holes out of the dot and radiative recombination ($T_1$ lifetime). Fits to the data using rate equation model of sequential electron and hole tunnelling (see refs \cite{Zecherle2010,Mueller2012} and supplementary information) are presented as solid lines and are in excellent agreement with the measured data, revealing an exciton lifetime of $175\pm5 \, ps$.  In figure 1d the spin-selectivity of the conditional absorption $X-2X$ as well as the stimulated emission $X-cgs$ remain unchanged for all time delays. In strong contrast, when exciting with circularly polarised light (figure 1e) we observe antiphased oscillations for co- and cross-polarised pump and probe pulses. This observation can be readily understood since initialising the exciton spin with a circularly polarised pulse creates an exciton spin state on the equatorial plane of the Bloch sphere.  This is a superposition of the energy eigenstates, whereupon the Bloch vector precesses about the z-axis.  The anti-phased oscillations observed in fig. 1e correspond to the projection of the exciton spin onto $R$ and $L$. For the data presented in fig. 1d, where we initialised an exciton spin into the $H$ state, the Pauli spin-blockade of the $X-2X$ transition and spin-selectivity of the stimulated emission remain the same for all time delays reflects the fact that the exciton spin decay time $T_1$ is much longer than the exciton lifetime. Clearly, the conditional absorption of $X-2X$ in fig. 1e repeatedly approaches zero and the envelope of the amplitude as obtained from fig. 1d for all time delays reflects the fact that the precession is \emph{fully coherent} with $T_2^*$ also much longer than the exciton lifetime. Thereby, the measured precession period of $T^{e-h}_{ex}=153\pm 1 \,ps$ indicates a fine structure splitting of $27 \, \mu eV$. In addition, the combined fidelity of initialisation and readout is estimated to be $\sim 97 \%$, limited only by the $\sim80$ fA noise in the photocurrent readout. As a key result, these results demonstrate that by probing the stimulated emission of the $cgs-X$ transition it is possible to read-out the spin projection in a fully resonant configuration of pump and probe.

\begin{figure}
\includegraphics[width=0.5\columnwidth]{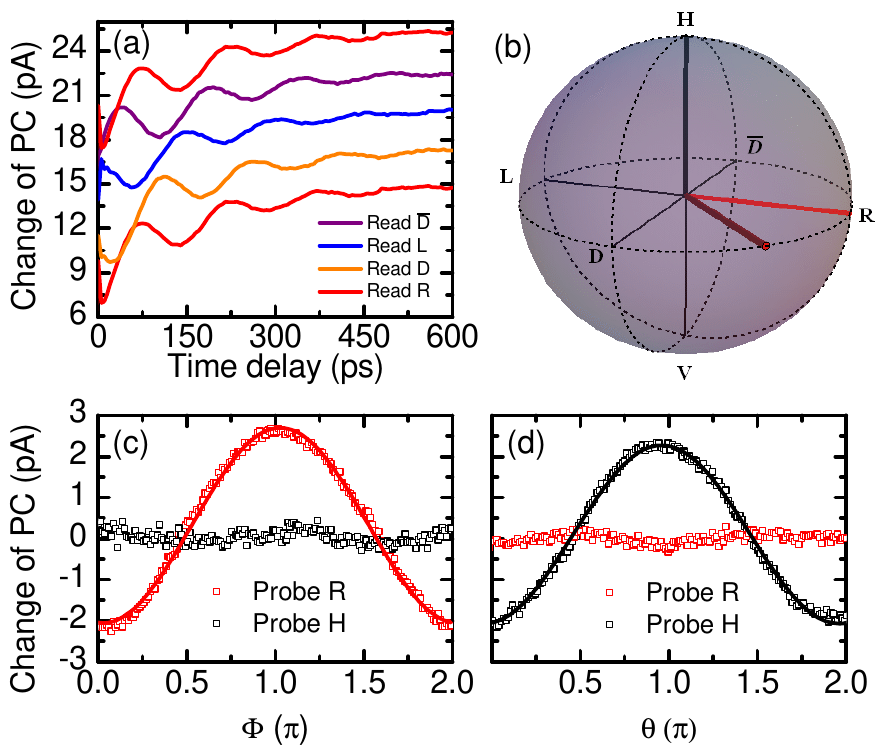}
\caption{\label{fig:Figure2}
(Color online) (a) Temporal evolution of $\Delta I$ for pumping excitons with R polarisation and probing the different projections R, D, L and $\bar{D}$. (b) Schematic illustration of the exciton spin on the Bloch sphere. (c) Change of PC for a fixed time delay of $38ps$, R (H) readout projection in red (black) for varying the initialisation angle $\phi = 0-2\pi$ $\theta = \pi/2$. (d) Same as (c) but for initialisation angles $\phi = 0$ $\theta = 0-\pi$.}
\end{figure}

Next, we demonstrate that we can initialise and read out arbitrary exciton spin states. To demonstrate the readout projection along specific axis, we present in fig. 2a $\Delta I$ as a function of the time delay upon probing the $cgs-X$ transition with the polarisation of the pump pulse fixed to $R$ and different readout polarisations $R$, $\bar{D}$, $L$ and $D$ (c.f. fig. 2b). Clearly extremely similar decaying oscillations are observed which are successively phase shifted by $T^{e-h}_{ex}/4=38\, ps$, reflecting the harmonic precession of the Bloch vector around the z-axis due to e-h exchange coupling. In order to demonstrate the arbitrary initialisation, we present in figs. 2c and 2d $\Delta I$ for a fixed time delay between pump and probe of $T^{e-h}_{ex}/4$ and two different readout projections $R$ and $H$ in red and black respectively. Thereby, the angle $\phi$ of the pump polarisation is varied continuously along the equatorial plane ($\theta = \pi/2$). As the figure shows, the projection $R$ (red curve) follows a sinusoidal dependence (solid line - fig. 2c). In contrast, the projection on $H$ (black curve) remains constant indicating a near perfect initialisation of the Bloch vector in the equatorial plane. In figure 2d the same projections are measured for varying the initialisation angle $\theta$ along a vertical cut through the Bloch sphere ($\phi = 0$). Here, the projection of $H$ follows a sinusoidal line (solid line - fig. 2d), while the projection on $R$ stays constant. Thereby, the errors of the fits indicate a combined fidelity of the initialisation and readout $ \geq 97 \%$ mainly limited by the 80fA readout noise.

\begin{figure}
\includegraphics[width=0.5\columnwidth]{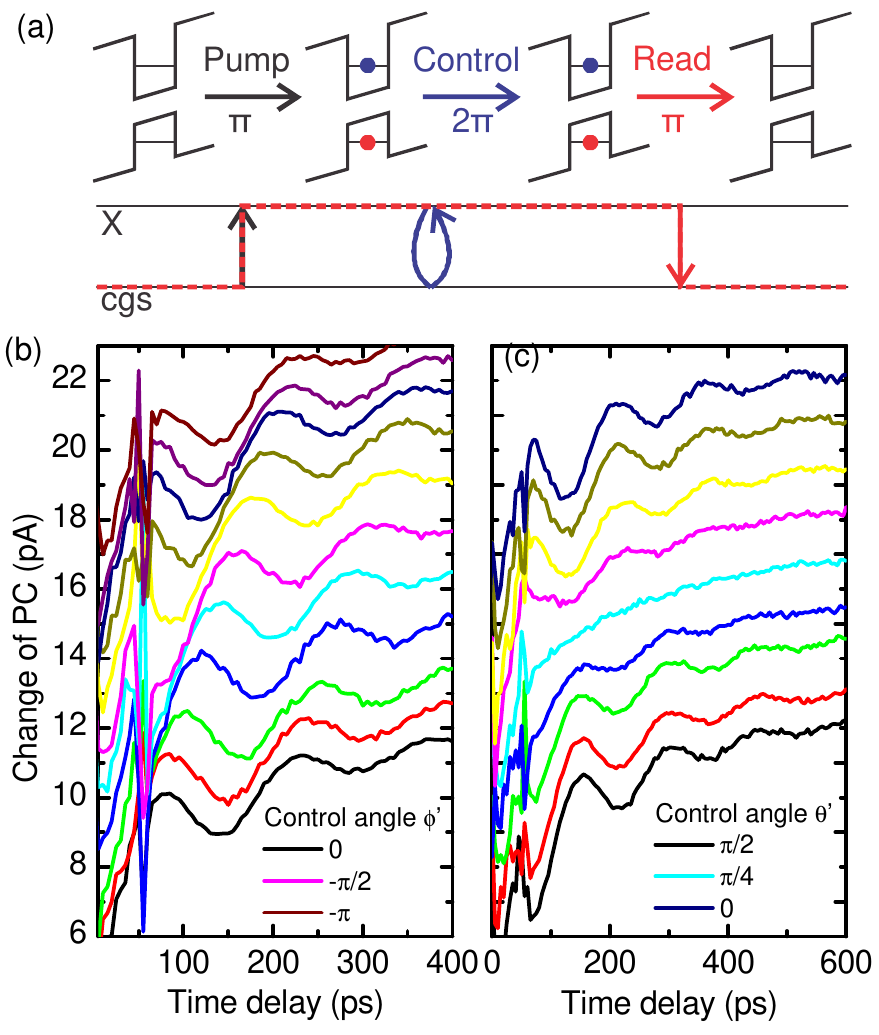}
\caption{\label{fig:Figure3}
(Color online) (a) Schematic illustration of the experiment. (b-c) Fully resonant coherent optical control for control and readout pulses tuned to $cgs-X$. The control angles are varied in (b) over the range $\phi' = 0-\pi$ $\theta' = \pi/2$ and in (c) from  $\phi' = 0$ $\theta' = 0-\pi/2$.}
\end{figure}

We continue to demonstrate ultrafast, high-fidelity and arbitrary optical control of the exciton spin wavefunction using the measurement sequence depicted schematically in fig. 3a. Our approach is based on the use of a precisely polarised $2\pi$-pulse tuned to be \emph{fully-resonant} with the $cgs-X$ transition\cite{non-resonant} to perform geometric phase control \cite{Economou2006,Economou2007,Takagahara2010,Poem2011}. 
Here, as the control pulse interacts with the system it coherently mixes the $\ket{cgs}$ and $\ket{X}$ states before returning the population fully to $\ket{X}$ as the pulse switches off. However, the phase accumulated by the exciton wavefunction during the presence of the control pulse is uniquely defined by the polarisation state of the control pulse - its direction on the Poincar\'e sphere\cite{Takagahara2010}. By setting the polarisation to a direction defined by the angles $\phi'$ and $\theta'$, with respect to initial orientation of the Bloch vector, the result of the control pulse is a rotation of the exciton spin Bloch vector by angles of $2\phi'$ and $2\theta'$ along the polar and azimuthal directions, respectively \cite{Economou2006,Economou2007,Takagahara2010}. Experimentally, we have tested these ideas by initialising the exciton spin on the equatorial plane using a $R$-circularly polarised $\pi$-pulse resonant with the $cgs-X$ transition. Following a time delay of $50 \, ps$, during which the phase of the spin wavefunction evolves freely, we apply the $2 \pi$-control laser pulse and finally read out the spin projection fully resonantly via the spin-selectivity of the $X \rightarrow cgs$ stimulated emission signal induced by a third, $R$-circularly polarised $\pi$-pulse.  The results of these investigations are presented in fig. 3 for spin rotations around two orthogonal axes - $\phi'$ and $\theta'$.

In fig. 3b we plot $\Delta I$ for different control polarisations $\phi'= 0-\pi$ and $\theta'=0$. Clearly, the oscillations recorded for different control pulse angles $\phi'$ have the same amplitude but are controllably phase-shifted by $2\cdot \phi'$, as expected. This demonstrates arbitrary rotations of the exciton spin Bloch vector throughout the equatorial plane.  In fig. 3c, we present similar experiments using control pulses which have the polarisation $\phi'=0$ and different angles $\theta' =0-\pi$. Clearly, the phase of oscillations induced by the e-h exchange coupling remain the same while the amplitude of the oscillations changes. Notably, for $\theta'=\pi/4$ the exciton spin is rotated to the top of the Bloch sphere and the amplitude of the oscillation vanishes since the system is rotated into the $H$ eigenstate.  For larger rotation angles $\theta'$, the exciton spin is rotated beyond the $z$-axis of the Bloch sphere, such that the amplitude of the oscillations changes its sign, as expected (fig. 3c). Similar measurements performed for arbitrary initialisation directions of the exciton spin and arbitrary rotations (not presented) revealed qualitatively similar levels of control.

The demonstration of arbitrary spin rotations about two orthogonal axes demonstrates arbitrary coherent control of the spin wavefunction over timescales as fast as $5 \, ps$ using a single, resonant optical pulse. It is important to note that these experiments demonstrate universal, all-optical exciton spin initialisation, control and readout over picosecond timescales. Moreover, as discussed in the supplementary information the amplitude of the oscillations remain practically unaffected by the control pulse, indicating that the control can be performed with very high fidelity.  The fully resonant preparation, control and readout scheme presented here is much more convenient than complex multi-colour experiments, especially when considering systems consisting of a number of interacting exciton qubits.  

\begin{figure}
\includegraphics[width=0.5\columnwidth]{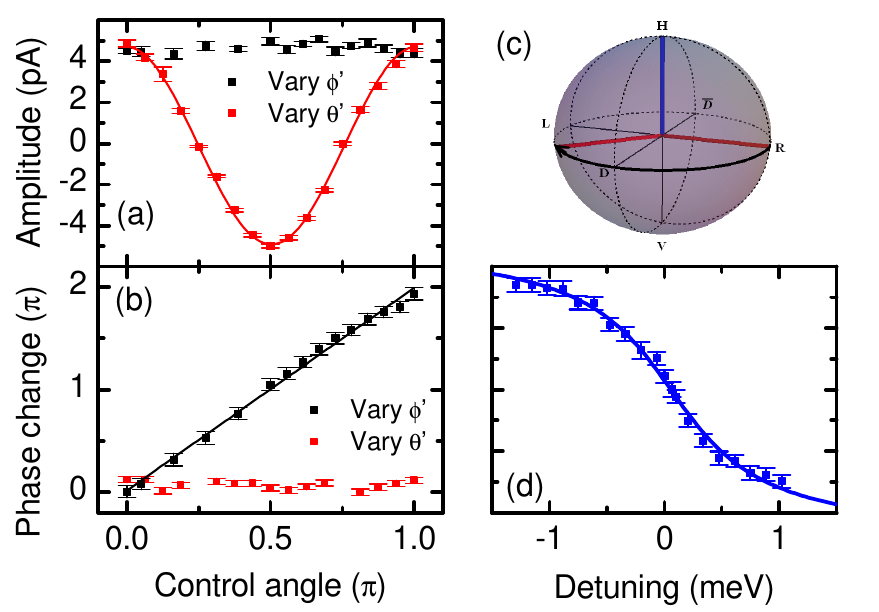}
\caption{\label{fig:Figure4}
(Color online) (a-b) Quantitative analysis of the data presented in figure 3(c-d): (a) Amplitudes and (b) phase change as a function of the control pulse angles $\phi'$ ($\theta'$) in black (red). (c-d) Effect of varying the control pulse detuning for the configuration illustrated in (c)}
\end{figure}

For a quantitative analysis, we fitted the data presented in fig. 3 using
\begin{equation}
\begin{split}
\Delta I(t) = O - A [\beta exp(-(\Gamma_e+\Gamma_r)t)-\alpha exp(-\Gamma_h t)] \\ - (A_1+A_2\cdot cos(t/T\cdot \pi-\omega_0)^2) exp(-(\Gamma_e + \Gamma_r)t)
\end{split}
\end{equation}

,where $\Gamma_{e(h)}$ is the electron(hole) tunnelling escape rate from the dot, $\Gamma_r\sim 1$ GHz the radiative recombination rate, $T$ the oscillation period due to e-h exchange and $\omega_0$ is the phase of the oscillation (see supplementary information). The amplitudes $A_1$ and $A_2$ denote the perpendicular and parallel amplitudes of the spin projection onto the polarisation of the readout pulse, quantities that are plotted in fig. 4a as a function of $\phi'$ (black points) and $\theta'$ (red points). Clearly the amplitudes for varying $\phi'$ remain the same within the experimental error ($\pm 200 fA$) while the amplitudes for varying $\theta'$ are very well represented by a cosine function. Thereby, small variations of the amplitudes appear to be random and, therefore, are most likely to result from laser power and polarisation fluctuations and drifts in the experimental setup and are not gating errors \cite{Takagahara2010}. The phase shifts as a function of the control pulse angles are presented in fig. 4b as a function $\phi'$ ($\theta'$) in black (red). Clearly, for varying $\phi'$, the phases vary with a slope very close to 2, as expected for geometric phase control\cite{Economou2007}, while for a variation of $\theta'$ the phases remain constant and shift by $\pi$ for control angles larger than $\pi/4$. To ensure that the small variations of the amplitude result from power variations and drifts and not from a reduced visibility of the fringes we repeated the experiment presented in fig. 3c but using non-resonant readout pulses tuned to the $X-2X$ transition. These experiments (not shown - see supplementary data) exhibited results very similar to the data shown in fig. 4, a comparable fidelity of $>96\%$ being observed, in good agreement with the values obtained here.

Finally, we explored the influence of energetically detuning the control pulse from the $cgs-X$ transition, since such detuned pulses have the potential to achieve phase control and further improve gating fidelity\cite{Benny2011,Takagahara2010} . The configuration of this experiment is schematically illustrated in figure 4c. For exciton spins at $\theta=\pi/2$ and $\phi=0$ we apply control pulses with a polarisation of $H$. Without detuning, the control pulse rotates the spin to $\phi=\pi$. However, the detuning of the control pulse energy from the $cgs-X$ transition determines the rotation angle about the z axis whereby the phase shift is given by
\begin{equation}
\delta = \pi-2 arctan\left(\frac{\Delta}{\sigma}\right)
\end{equation}
for secant pulses with a bandwidth $\sigma$ and a detuning $\Delta$ \cite{Economou2006,Economou2007, Kodriano2012}. The results of these experiments are presented in fig. 4d that shows the phase shift as a function of the control pulse detuning.  A fit to the data with eqn 1 is presented as a solid line and produces excellent agreement using a pulse bandwidth of $0.6\pm0.1 \, meV$ in excellent accord with the measured control pulse bandwidth of $0.5\pm0.1 \, meV$ and measured pulse duration of $4\pm1 \, ps$ using autocorrelation.

In summary, we have demonstrated the arbitrary initialisation, full coherent control and readout of a single exciton spin in an InGaAs quantum dot using ps pulses. We have presented a direct mapping of the polarisation of a resonant laser pulse to the exciton spin with a combined initialisation and readout fidelity $>97\%$. We have shown that the spin can be read out either by the spin selective absorption of the $X-2X$ transition as well as the spin-selective stimulated emission of $X\rightarrow cgs$. In addition, we have presented the high-fidelity full coherent optical control using ps-pulses that are also resonant with the $cgs-X$ transition. 

\textit{We gratefully acknowledge financial support from the DFG via SFB-631, the Nanosystems Initiative Munich (NIM) and the EU via the integrated project SOLID. GA thanks the TUM Institute for Advanced Study for support.}

\section{Sample structure}
The samples investigated consist of a single layer of low density ($\sim5 \mu m^{-1}$) self-assembled InGaAs quantum dots grown via molecular beam epitaxy embedded within the intrinsic region of a $300nm$ thick n-type GaAs Schottky photodiode. The layer structure and resulting band diagram are schematically illustrated in figures S1a and b, respectively. Such devices facilitate complementary photocurrent absorption (PC) and photoluminescence emission (PL) measurements as a function of the internal electric field ($F$)\cite{Fry2000, Findeis2001}. To spatially address single quantum dots for optical investigation and make it easier to re-locate them for systematic studies $\sim 1 \, \mu m$ sized apertures are fabricated in the opaque 200nm thick gold layer on the top contact. A scanning electron microscopy image of a typical aperture is presented in figure S1c. These apertures are fabricated by the deposition of polystyrene nanospheres before the deposition of the gold layer. The polystyrene nanospheres act as shadow masks during the deposition of the gold layer and are mechanically removed after Au deposition to produce circular apertures through which optical spectroscopy measurements can be performed.

\begin{figure}
\includegraphics[width=0.5\columnwidth]{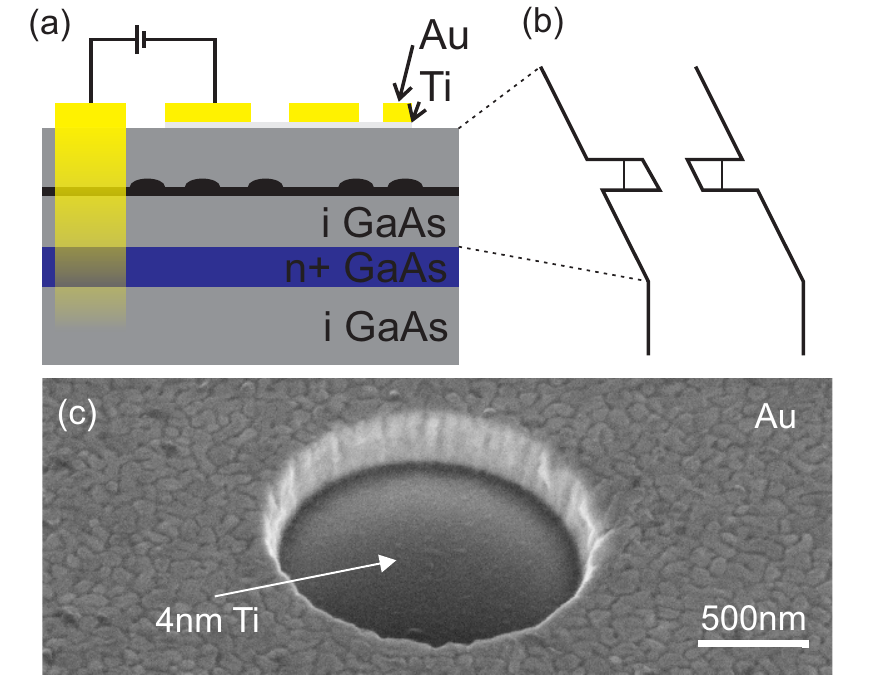}
\caption{\label{fig:FigureS0}
(Color online) (a-b)Schematic illustration of the sample structure (a) and resulting band structure (b). (c) Scanning electron microscopy image of an aperture}
\end{figure}

\section{Measurement techniques}
The primary measurement technique used in this paper is ultrafast pump-probe spectroscopy. The setup used for the pump-probe experiments with photocurrent readout is presented schematically in figure S2. Starting with a $\sim 100 \,fs$ duration pulse from a tunable Ti:Sa laser (figure S1a left) three independently tunable pulse trains (pump, control and the probe pulses) are derived using a balanced set of three 4f pulse-shapers (figure S1a bottom). The relative time delay between the three pulse trains is precisely controlled using a delay line (figure S1a center) that provides a temporal relative tuning range from $-300 \, ps \,- \,+1 \,ns$. The operating  principle of the 4f pulse-shaper is illustrated in figure S1b and ref. \cite{Ramsay08}. The incident beam (left) is spectrally dispersed using a $1200 \, l/mm$ ruled grating and made parallel using an $f = 500 \, mm$ lens. A tunable slit positioned on a motorised linear stage (center) filters the light in the Fourier plane to transmit spectrally narrow pulses out of the broadband input. The power of the three pulses can be individually tuned using variable neutral density filters. The polarisation of the three pulses is individually controlled using the combination of a $\lambda/2$ plate, a thin film linear polariser and liquid crystal retarders. Measurements of the performance of our pulse shapers are presented in ref \cite{Mueller2012_3}.The pulses are sent to a low-temperature confocal microscope using optical fibres where they are superimposed co-linearly and focused onto the sample. Thereby, the low-temperature microscope consists of a sample stick filled with He-exchange gas inside a liquid helium dewar and the sample is moved using attocube piezo stepper motors. The photocurrent $I$ induced in the sample is measured using a low-noise Ithaco model 1211 current-voltage converter and an Agilent 34411A digital multimeter. A programmable voltage source (Keithley model 2400) connected in series with the amplifier is used to apply a gate voltage to the sample and, thus, control the internal electric field in the intrinsic region of the devices into which the QDs are embedded. Blocking and unblocking the pump beam reveals the pump induced change of the probe induced photocurrent $\Delta I$. The quantities $I$ and $\Delta I$ can be interpreted as the absorption of the QD nanostructure and its \textit{pump induced change}, respectively \cite{Zecherle2010}. In this representation, a positive value of $\Delta I$ corresponds to pump induced absorption whereas negative $\Delta I$ corresponds to bleaching.

\begin{figure}
\includegraphics[width=0.5\columnwidth]{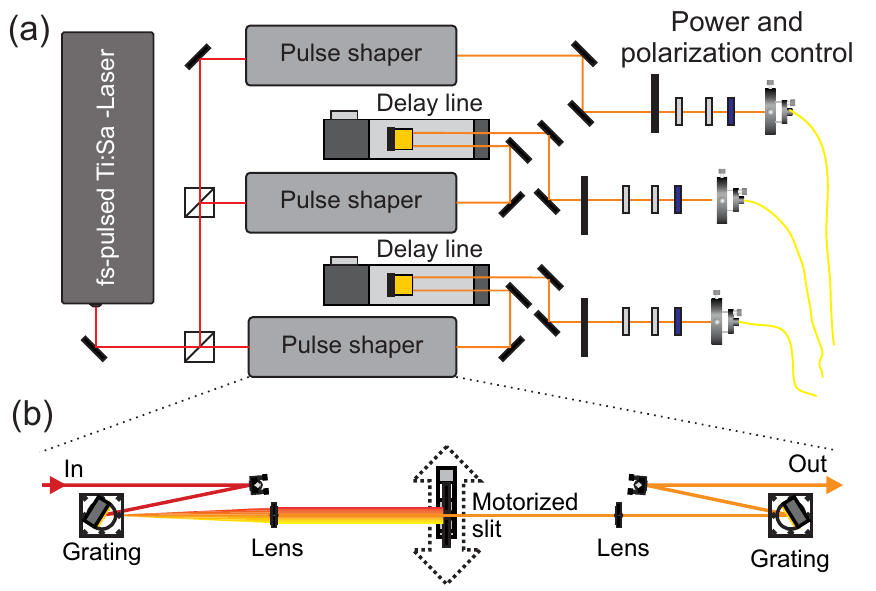}
\caption{\label{fig:FigureS1}
(Color online) Schematic illustration of the setup for ultrafast pump-control-probe spectroscopy}
\end{figure}

\section{Rate equation model}
In order to fit the measurements of $\Delta I$ as a function of the time delay between pump and probe we use a rate equation model of sequential electron and hole tunnelling \cite{Zecherle2010, Mueller2012, Mueller2012-1, Mueller2012_3}. In this model, at delay times of $t_d =0 ps$ the pump pulse generates the initial occupation $N_{X}(t=0)=N_0$ of the $X$ state and $N_{cgs}(t=0)=1-N_0$ of the crystal ground state (cgs). Subsequently, electrons and holes tunnel out with the independent rates $\Gamma_e$ and $\Gamma_h$, respectively. Thereby, the tunnelling time of the electron ($t_e = 1/ \Gamma_e$) with its smaller effective mass is much shorter than that of the hole ($t_h = 1 / \Gamma_h$) \cite{Fry2000, Oulton2002}. Therefore, the model takes into account the intermediate level where the QD is occupied with a single hole $N_h$. Taking also into account the radiative recombination of the exciton with the rate $\Gamma_r$ the set of differential equations that describe the rate equation model can be written as:

\begin{equation}
\frac{d}{dt}\left( \begin{matrix} N_X(t) \\ N_h(t) \\ N_{cgs}(t) \end{matrix} \right) =
\left( \begin{matrix} -(\Gamma_e + \Gamma_r) & 0 & 0 \\ \Gamma_e & -\Gamma_h & 0 \\ \Gamma_r & \Gamma_h & 0 \end{matrix} \right) \cdot
\left( \begin{matrix} N_X(t) \\ N_h(t) \\ N_{cgs}(t) \end{matrix} \right)
\end{equation}

Then, the analytic solution of the rate equations that describe the time dependent populations under these initial conditions are
\begin{equation}
\begin{split}
N_{X}(t) = N_0 exp(-(\Gamma_e+\Gamma_r)t) \\
N_{h}(t) = N_0 \alpha [exp(-(\Gamma_e+\Gamma_r)t) - exp(- \Gamma_h t)] \\
N_{cgs}(t) = 1- N_{X}(t)-N_{h}(t)
\end{split}
\end{equation}
with $\alpha = \frac{\Gamma_e}{\Gamma_h - \Gamma_e - \Gamma_r}$. Probing an optical transition results in a probe-induced photocurrent with an amplitude that is proportional to the difference of the occupations of the initial state and the final state. Since the final state of the $h^+ \rightarrow X^+$ and $X \rightarrow 2X$ transitions are not occupied before probing, the number of the positively charged QDs ($N_h$) and the X population ($N_{X}$) can be directly compared to the temporal scans probing the $h^+ \rightarrow X^+$ and $X \rightarrow 2X$ transitions, respectively. Probing the $cgs \rightarrow X$ transition results in a population inversion ($N_{cgs}-N_{X}$) and the fitting functions are:

\begin{equation}
I_{X-2X}(t) = O + A exp(-(\Gamma_e+\Gamma_r)t)
\end{equation}
\begin{equation}
I_{h-X^+}(t) = O + A \alpha [exp(-(\Gamma_e+\Gamma_r)t) - exp(- \Gamma_h t)]
\end{equation}
\begin{equation}
\begin{split}
I_{cgs-X}(t) = O - A [\beta exp(-(\Gamma_e+\Gamma_r)t)-\alpha exp(-\Gamma_h t)] \\ - A exp(-(\Gamma_e + \Gamma_r)t)
\end{split}
\end{equation}

with $\beta = \frac{\Gamma_h-\Gamma_r}{\Gamma_h-\Gamma_e-\Gamma_r}$. Finally, taking into account the spin selectivity of the conditional absorption $X \rightarrow 2X$ and stimulated emission $X \rightarrow cgs$ the fitting functions are:

\begin{equation}
I_{X-2X}(t) = O + (A_1+A_2\cdot cos(t/T\cdot \pi-\omega_0)^2) exp(-(\Gamma_e+\Gamma_r)t)
\end{equation}
\begin{equation}
\begin{split}
I_{cgs-X}(t) = O - A [\beta exp(-(\Gamma_e+\Gamma_r)t)-\alpha exp(-\Gamma_h t)] \\ - (A_1+A_2\cdot cos(t/T\cdot \pi-\omega_0)^2) exp(-(\Gamma_e + \Gamma_r)t)
\end{split}
\end{equation}

whereby $T$ denotes the oscillation period and $\omega_0$ the phase. The amplitudes $A_1$ and $A_2$ denote the amplitudes of the projection on the polarisation of the readout pulse and perpendicular to it.

\section{Estimation of fidelity}

\begin{figure}
\includegraphics[width=0.5\columnwidth]{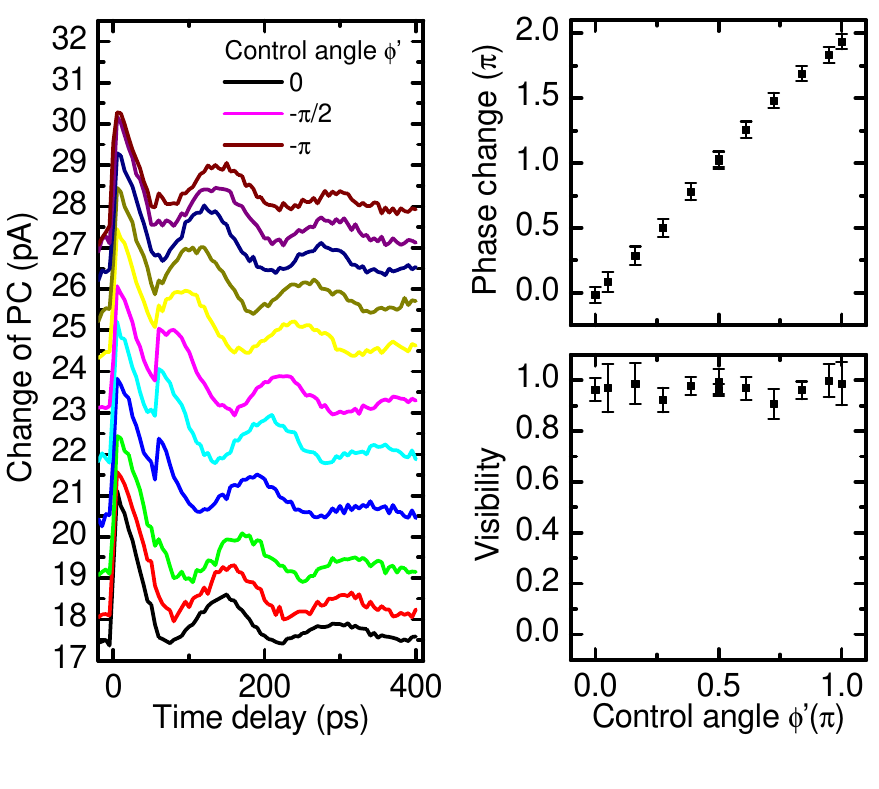}
\caption{\label{fig:FigureS2}
(Color online) Coherent optical control for control pulses in resonance with $cgs-X$ and readout pulses in resonance with $X-2X$. (a)Spectra for different control angles $\phi'$. (b)Phase shift and (c) visibility of the oscillations $A_2 / (|A_1| + |A_2|)$ from the fits as discussed above.}
\end{figure}

In figures 3 and 4 of the paper we performed the coherent control and readout for pulses in resonance with the $cgs-X$ transition. Therefore, the oscillating stimulated emission of the readout is superimposed with the exponential bleaching of the $cgs-X$ transition. In order to investigate the fidelity of the coherent control in more detail, we performed the same experiment as presented in figure 3b for readout pulses tuned to the $X-2X$ transition. The result of this experiment is presented in figure S3a. The figure shows $\Delta I$ for different control angles $\phi' = 0 - \pi$ and $\theta'=0$. Clearly oscillations that are fully modulated and shifted in phase are observed. For a quantitative analysis we fit the data with the rate equation model described above and present the phase shifts in figure S3b and the visibility of the oscillations $A_2 / (|A_1| + |A_2|)$ in figure S3c as a function of the control pulse polarisation angle $\phi'$. Clearly the phase shift (figure S3b) has a slope of two while the visibility of the oscillations (figure S3c) is close to 1 with no systematic variations. A statistical analysis results a visibility of $96 \pm 3 \%$.

\section{Control via the $X-2X$ transition}

\begin{figure}
\includegraphics[width=0.5\columnwidth]{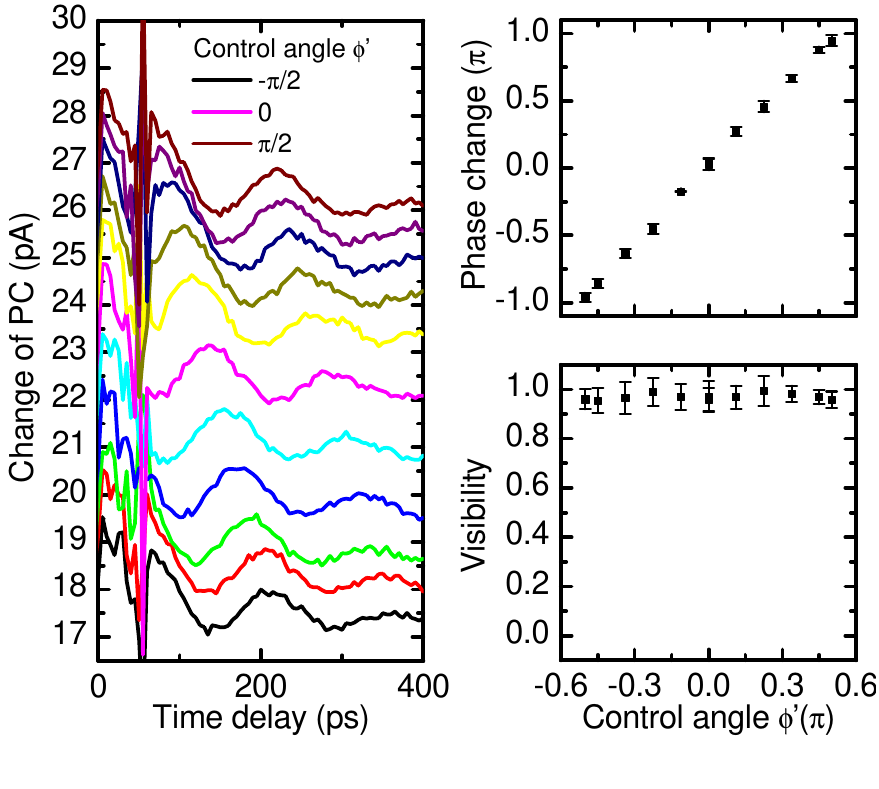}
\caption{\label{fig:FigureS3}
(Color online) Coherent optical control for control pulses and readout pulses in resonance with $X-2X$. (a)Spectra for different control angles $\phi'$. (b)Phaseshift and (c) visibility of the oscillations from fits with the rate equation model.}
\end{figure}

To demonstrate that the coherent optical control is also possible for control pulses tuned to the $X-2X$ transition we present the same experiment as presented in figure S3 in figure S4 for control pulses tuned to the $X-2X$ transition. Again, figure S4a shows $\Delta I$ for different control pulse polarisation angles $\phi' = 0 - \pi$ and $\theta'=0$ and fully modulated oscillations that are phase shifted are observed. The results of fits with the rate equation model described above are presented in figure S4b and c that show phase shifts in figure S4b and the visibility of the oscillations in figure S4c as a function of the control pulse polarisation angle $\phi'$. Again, the slope of the phase shift (figure S4b) is 2, demonstrating that coherent control is also possible for control pulses tuned to the $X-2X$ transition and and analysis of the visibility of the oscillations (figure S4c) results a value of $97 \pm 1.5 \%$.


\begin{thebibliography}{36}%
\makeatletter
\providecommand \@ifxundefined [1]{%
 \@ifx{#1\undefined}
}%
\providecommand \@ifnum [1]{%
 \ifnum #1\expandafter \@firstoftwo
 \else \expandafter \@secondoftwo
 \fi
}%
\providecommand \@ifx [1]{%
 \ifx #1\expandafter \@firstoftwo
 \else \expandafter \@secondoftwo
 \fi
}%
\providecommand \natexlab [1]{#1}%
\providecommand \enquote  [1]{``#1''}%
\providecommand \bibnamefont  [1]{#1}%
\providecommand \bibfnamefont [1]{#1}%
\providecommand \citenamefont [1]{#1}%
\providecommand \href@noop [0]{\@secondoftwo}%
\providecommand \href [0]{\begingroup \@sanitize@url \@href}%
\providecommand \@href[1]{\@@startlink{#1}\@@href}%
\providecommand \@@href[1]{\endgroup#1\@@endlink}%
\providecommand \@sanitize@url [0]{\catcode `\\12\catcode `\$12\catcode
  `\&12\catcode `\#12\catcode `\^12\catcode `\_12\catcode `\%12\relax}%
\providecommand \@@startlink[1]{}%
\providecommand \@@endlink[0]{}%
\providecommand \url  [0]{\begingroup\@sanitize@url \@url }%
\providecommand \@url [1]{\endgroup\@href {#1}{\urlprefix }}%
\providecommand \urlprefix  [0]{URL }%
\providecommand \Eprint [0]{\href }%
\providecommand \doibase [0]{http://dx.doi.org/}%
\providecommand \selectlanguage [0]{\@gobble}%
\providecommand \bibinfo  [0]{\@secondoftwo}%
\providecommand \bibfield  [0]{\@secondoftwo}%
\providecommand \translation [1]{[#1]}%
\providecommand \BibitemOpen [0]{}%
\providecommand \bibitemStop [0]{}%
\providecommand \bibitemNoStop [0]{.\EOS\space}%
\providecommand \EOS [0]{\spacefactor3000\relax}%
\providecommand \BibitemShut  [1]{\csname bibitem#1\endcsname}%
\let\auto@bib@innerbib\@empty
%</preamble>
\bibitem [{\citenamefont {Loss}\ and\ \citenamefont
  {DiVincenzo}(1998)}]{Loss1998}%
  \BibitemOpen
  \bibfield  {author} {\bibinfo {author} {\bibfnamefont {D.}~\bibnamefont
  {Loss}}\ and\ \bibinfo {author} {\bibfnamefont {D.~P.}\ \bibnamefont
  {DiVincenzo}},\ }\href {\doibase 10.1103/PhysRevA.57.120} {\bibfield
  {journal} {\bibinfo  {journal} {Phys. Rev. A}\ }\textbf {\bibinfo {volume}
  {57}},\ \bibinfo {pages} {120} (\bibinfo {year} {1998})}\BibitemShut
  {NoStop}%
\bibitem [{\citenamefont {Imamoglu}\ \emph {et~al.}(1999)\citenamefont
  {Imamoglu}, \citenamefont {Awschalom}, \citenamefont {Burkard}, \citenamefont
  {DiVincenzo}, \citenamefont {Loss}, \citenamefont {Sherwin},\ and\
  \citenamefont {Small}}]{Imamoglu1999}%
  \BibitemOpen
  \bibfield  {author} {\bibinfo {author} {\bibfnamefont {A.}~\bibnamefont
  {Imamoglu}}, \bibinfo {author} {\bibfnamefont {D.~D.}\ \bibnamefont
  {Awschalom}}, \bibinfo {author} {\bibfnamefont {G.}~\bibnamefont {Burkard}},
  \bibinfo {author} {\bibfnamefont {D.~P.}\ \bibnamefont {DiVincenzo}},
  \bibinfo {author} {\bibfnamefont {D.}~\bibnamefont {Loss}}, \bibinfo {author}
  {\bibfnamefont {M.}~\bibnamefont {Sherwin}}, \ and\ \bibinfo {author}
  {\bibfnamefont {A.}~\bibnamefont {Small}},\ }\href {\doibase
  10.1103/PhysRevLett.83.4204} {\bibfield  {journal} {\bibinfo  {journal}
  {Phys. Rev. Lett.}\ }\textbf {\bibinfo {volume} {83}},\ \bibinfo {pages}
  {4204} (\bibinfo {year} {1999})}\BibitemShut {NoStop}%
\bibitem [{\citenamefont {Atatüre}\ \emph {et~al.}(2006)\citenamefont
  {Atatüre}, \citenamefont {Dreiser}, \citenamefont {Badolato}, \citenamefont
  {Högele}, \citenamefont {Karrai},\ and\ \citenamefont {Imamoglu}}]{Mete2006}%
  \BibitemOpen
  \bibfield  {author} {\bibinfo {author} {\bibfnamefont {M.}~\bibnamefont
  {Atatüre}}, \bibinfo {author} {\bibfnamefont {J.}~\bibnamefont {Dreiser}},
  \bibinfo {author} {\bibfnamefont {A.}~\bibnamefont {Badolato}}, \bibinfo
  {author} {\bibfnamefont {A.}~\bibnamefont {Högele}}, \bibinfo {author}
  {\bibfnamefont {K.}~\bibnamefont {Karrai}}, \ and\ \bibinfo {author}
  {\bibfnamefont {A.}~\bibnamefont {Imamoglu}},\ }\href {\doibase
  10.1126/science.1126074} {\bibfield  {journal} {\bibinfo  {journal}
  {Science}\ }\textbf {\bibinfo {volume} {312}},\ \bibinfo {pages} {551}
  (\bibinfo {year} {2006})},\ \Eprint
  {http://arxiv.org/abs/http://www.sciencemag.org/content/312/5773/551.full.pdf}
  {http://www.sciencemag.org/content/312/5773/551.full.pdf} \BibitemShut
  {NoStop}%
\bibitem [{\citenamefont {Kim}\ \emph {et~al.}(2011)\citenamefont {Kim},
  \citenamefont {Carter}, \citenamefont {Greilich}, \citenamefont {Bracker},\
  and\ \citenamefont {Gammon}}]{Kim2010}%
  \BibitemOpen
  \bibfield  {author} {\bibinfo {author} {\bibfnamefont {D.}~\bibnamefont
  {Kim}}, \bibinfo {author} {\bibfnamefont {S.~G.}\ \bibnamefont {Carter}},
  \bibinfo {author} {\bibfnamefont {A.}~\bibnamefont {Greilich}}, \bibinfo
  {author} {\bibfnamefont {A.~S.}\ \bibnamefont {Bracker}}, \ and\ \bibinfo
  {author} {\bibfnamefont {D.}~\bibnamefont {Gammon}},\ }\href {\doibase
  10.1038/NPHYS1863} {\bibfield  {journal} {\bibinfo  {journal} {Nature
  Physics}\ }\textbf {\bibinfo {volume} {7}},\ \bibinfo {pages} {223} (\bibinfo
  {year} {2011})}\BibitemShut {NoStop}%
\bibitem [{\citenamefont {Greilich}\ \emph {et~al.}(2011)\citenamefont
  {Greilich}, \citenamefont {Carter}, \citenamefont {Kim}, \citenamefont
  {Bracker},\ and\ \citenamefont {Gammon}}]{Greilich2011}%
  \BibitemOpen
  \bibfield  {author} {\bibinfo {author} {\bibfnamefont {A.}~\bibnamefont
  {Greilich}}, \bibinfo {author} {\bibfnamefont {S.~G.}\ \bibnamefont
  {Carter}}, \bibinfo {author} {\bibfnamefont {D.}~\bibnamefont {Kim}},
  \bibinfo {author} {\bibfnamefont {A.~S.}\ \bibnamefont {Bracker}}, \ and\
  \bibinfo {author} {\bibfnamefont {D.}~\bibnamefont {Gammon}},\ }\href
  {\doibase 10.1038/NPHOTON.2011.237} {\bibfield  {journal} {\bibinfo
  {journal} {Nature Photonics}\ }\textbf {\bibinfo {volume} {5}},\ \bibinfo
  {pages} {703} (\bibinfo {year} {2011})}\BibitemShut {NoStop}%
\bibitem [{\citenamefont {Weiss}\ \emph {et~al.}(2012)\citenamefont {Weiss},
  \citenamefont {Elzerman}, \citenamefont {Delley}, \citenamefont
  {Miguel-Sanchez},\ and\ \citenamefont {Imamo\ifmmode~\breve{g}\else
  \u{g}\fi{}lu}}]{Weiss2012}%
  \BibitemOpen
  \bibfield  {author} {\bibinfo {author} {\bibfnamefont {K.~M.}\ \bibnamefont
  {Weiss}}, \bibinfo {author} {\bibfnamefont {J.~M.}\ \bibnamefont {Elzerman}},
  \bibinfo {author} {\bibfnamefont {Y.~L.}\ \bibnamefont {Delley}}, \bibinfo
  {author} {\bibfnamefont {J.}~\bibnamefont {Miguel-Sanchez}}, \ and\ \bibinfo
  {author} {\bibfnamefont {A.}~\bibnamefont {Imamo\ifmmode~\breve{g}\else
  \u{g}\fi{}lu}},\ }\href {\doibase 10.1103/PhysRevLett.109.107401} {\bibfield
  {journal} {\bibinfo  {journal} {Phys. Rev. Lett.}\ }\textbf {\bibinfo
  {volume} {109}},\ \bibinfo {pages} {107401} (\bibinfo {year}
  {2012})}\BibitemShut {NoStop}%
\bibitem [{\citenamefont {Kroutvar}\ \emph {et~al.}(2004)\citenamefont
  {Kroutvar}, \citenamefont {Ducommun}, \citenamefont {Heiss}, \citenamefont
  {Bichler}, \citenamefont {Schuh}, \citenamefont {Abstreiter},\ and\
  \citenamefont {Finley}}]{Kroutvar2004}%
  \BibitemOpen
  \bibfield  {author} {\bibinfo {author} {\bibfnamefont {M.}~\bibnamefont
  {Kroutvar}}, \bibinfo {author} {\bibfnamefont {Y.}~\bibnamefont {Ducommun}},
  \bibinfo {author} {\bibfnamefont {D.}~\bibnamefont {Heiss}}, \bibinfo
  {author} {\bibfnamefont {M.}~\bibnamefont {Bichler}}, \bibinfo {author}
  {\bibfnamefont {D.}~\bibnamefont {Schuh}}, \bibinfo {author} {\bibfnamefont
  {G.}~\bibnamefont {Abstreiter}}, \ and\ \bibinfo {author} {\bibfnamefont
  {J.}~\bibnamefont {Finley}},\ }\href {\doibase 10.1038/nature03008}
  {\bibfield  {journal} {\bibinfo  {journal} {Nature}\ }\textbf {\bibinfo
  {volume} {432}},\ \bibinfo {pages} {81} (\bibinfo {year} {2004})}\BibitemShut
  {NoStop}%
\bibitem [{\citenamefont {Heiss}\ \emph {et~al.}(2007)\citenamefont {Heiss},
  \citenamefont {Schaeck}, \citenamefont {Huebl}, \citenamefont {Bichler},
  \citenamefont {Abstreiter}, \citenamefont {Finley}, \citenamefont {Bulaev},\
  and\ \citenamefont {Loss}}]{Heiss2007}%
  \BibitemOpen
  \bibfield  {author} {\bibinfo {author} {\bibfnamefont {D.}~\bibnamefont
  {Heiss}}, \bibinfo {author} {\bibfnamefont {S.}~\bibnamefont {Schaeck}},
  \bibinfo {author} {\bibfnamefont {H.}~\bibnamefont {Huebl}}, \bibinfo
  {author} {\bibfnamefont {M.}~\bibnamefont {Bichler}}, \bibinfo {author}
  {\bibfnamefont {G.}~\bibnamefont {Abstreiter}}, \bibinfo {author}
  {\bibfnamefont {J.~J.}\ \bibnamefont {Finley}}, \bibinfo {author}
  {\bibfnamefont {D.~V.}\ \bibnamefont {Bulaev}}, \ and\ \bibinfo {author}
  {\bibfnamefont {D.}~\bibnamefont {Loss}},\ }\href {\doibase
  10.1103/PhysRevB.76.241306} {\bibfield  {journal} {\bibinfo  {journal} {Phys.
  Rev. B}\ }\textbf {\bibinfo {volume} {76}},\ \bibinfo {pages} {241306}
  (\bibinfo {year} {2007})}\BibitemShut {NoStop}%
\bibitem [{\citenamefont {Heiss}\ \emph {et~al.}(2008)\citenamefont {Heiss},
  \citenamefont {Jovanov}, \citenamefont {Bichler}, \citenamefont
  {Abstreiter},\ and\ \citenamefont {Finley}}]{Heiss2008}%
  \BibitemOpen
  \bibfield  {author} {\bibinfo {author} {\bibfnamefont {D.}~\bibnamefont
  {Heiss}}, \bibinfo {author} {\bibfnamefont {V.}~\bibnamefont {Jovanov}},
  \bibinfo {author} {\bibfnamefont {M.}~\bibnamefont {Bichler}}, \bibinfo
  {author} {\bibfnamefont {G.}~\bibnamefont {Abstreiter}}, \ and\ \bibinfo
  {author} {\bibfnamefont {J.~J.}\ \bibnamefont {Finley}},\ }\href {\doibase
  10.1103/PhysRevB.77.235442} {\bibfield  {journal} {\bibinfo  {journal} {Phys.
  Rev. B}\ }\textbf {\bibinfo {volume} {77}},\ \bibinfo {pages} {235442}
  (\bibinfo {year} {2008})}\BibitemShut {NoStop}%
\bibitem [{\citenamefont {Godden}\ \emph {et~al.}(2010)\citenamefont {Godden},
  \citenamefont {Boyle}, \citenamefont {Ramsay}, \citenamefont {Fox},\ and\
  \citenamefont {Skolnick}}]{Godden2010}%
  \BibitemOpen
  \bibfield  {author} {\bibinfo {author} {\bibfnamefont {T.~M.}\ \bibnamefont
  {Godden}}, \bibinfo {author} {\bibfnamefont {S.~J.}\ \bibnamefont {Boyle}},
  \bibinfo {author} {\bibfnamefont {A.~J.}\ \bibnamefont {Ramsay}}, \bibinfo
  {author} {\bibfnamefont {A.~M.}\ \bibnamefont {Fox}}, \ and\ \bibinfo
  {author} {\bibfnamefont {M.~S.}\ \bibnamefont {Skolnick}},\ }\href {\doibase
  10.1063/1.3476353} {\bibfield  {journal} {\bibinfo  {journal} {Applied
  Physics Letters}\ }\textbf {\bibinfo {volume} {97}},\ \bibinfo {eid} {061113}
  (\bibinfo {year} {2010})}\BibitemShut {NoStop}%
\bibitem [{\citenamefont {Godden}\ \emph {et~al.}(2012)\citenamefont {Godden},
  \citenamefont {Quilter}, \citenamefont {Ramsay}, \citenamefont {Wu},
  \citenamefont {Brereton}, \citenamefont {Boyle}, \citenamefont {Luxmoore},
  \citenamefont {Puebla-Nunez}, \citenamefont {Fox},\ and\ \citenamefont
  {Skolnick}}]{Godden2012}%
  \BibitemOpen
  \bibfield  {author} {\bibinfo {author} {\bibfnamefont {T.~M.}\ \bibnamefont
  {Godden}}, \bibinfo {author} {\bibfnamefont {J.~H.}\ \bibnamefont {Quilter}},
  \bibinfo {author} {\bibfnamefont {A.~J.}\ \bibnamefont {Ramsay}}, \bibinfo
  {author} {\bibfnamefont {Y.}~\bibnamefont {Wu}}, \bibinfo {author}
  {\bibfnamefont {P.}~\bibnamefont {Brereton}}, \bibinfo {author}
  {\bibfnamefont {S.~J.}\ \bibnamefont {Boyle}}, \bibinfo {author}
  {\bibfnamefont {I.~J.}\ \bibnamefont {Luxmoore}}, \bibinfo {author}
  {\bibfnamefont {J.}~\bibnamefont {Puebla-Nunez}}, \bibinfo {author}
  {\bibfnamefont {A.~M.}\ \bibnamefont {Fox}}, \ and\ \bibinfo {author}
  {\bibfnamefont {M.~S.}\ \bibnamefont {Skolnick}},\ }\href {\doibase
  10.1103/PhysRevLett.108.017402} {\bibfield  {journal} {\bibinfo  {journal}
  {Phys. Rev. Lett.}\ }\textbf {\bibinfo {volume} {108}},\ \bibinfo {pages}
  {017402} (\bibinfo {year} {2012})}\BibitemShut {NoStop}%
\bibitem [{\citenamefont {Ramsay}\ \emph {et~al.}(2008)\citenamefont {Ramsay},
  \citenamefont {Boyle}, \citenamefont {Kolodka}, \citenamefont {Oliveira},
  \citenamefont {Skiba-Szymanska}, \citenamefont {Liu}, \citenamefont
  {Hopkinson}, \citenamefont {Fox},\ and\ \citenamefont {Skolnick}}]{Ramsay08}%
  \BibitemOpen
  \bibfield  {author} {\bibinfo {author} {\bibfnamefont {A.~J.}\ \bibnamefont
  {Ramsay}}, \bibinfo {author} {\bibfnamefont {S.~J.}\ \bibnamefont {Boyle}},
  \bibinfo {author} {\bibfnamefont {R.~S.}\ \bibnamefont {Kolodka}}, \bibinfo
  {author} {\bibfnamefont {J.~B.~B.}\ \bibnamefont {Oliveira}}, \bibinfo
  {author} {\bibfnamefont {J.}~\bibnamefont {Skiba-Szymanska}}, \bibinfo
  {author} {\bibfnamefont {H.~Y.}\ \bibnamefont {Liu}}, \bibinfo {author}
  {\bibfnamefont {M.}~\bibnamefont {Hopkinson}}, \bibinfo {author}
  {\bibfnamefont {A.~M.}\ \bibnamefont {Fox}}, \ and\ \bibinfo {author}
  {\bibfnamefont {M.~S.}\ \bibnamefont {Skolnick}},\ }\href {\doibase
  10.1103/PhysRevLett.100.197401} {\bibfield  {journal} {\bibinfo  {journal}
  {Phys. Rev. Lett.}\ }\textbf {\bibinfo {volume} {100}} (\bibinfo {year}
  {2008}),\ 10.1103/PhysRevLett.100.197401}\BibitemShut {NoStop}%
\bibitem [{\citenamefont {Muller}\ \emph {et~al.}(2012)\citenamefont {Muller},
  \citenamefont {Bechtold}, \citenamefont {Ruppert}, \citenamefont {Hautmann},
  \citenamefont {Wildmann}, \citenamefont {Kaldewey}, \citenamefont {Bichler},
  \citenamefont {Krenner}, \citenamefont {Abstreiter}, \citenamefont {Betz},\
  and\ \citenamefont {Finley}}]{Mueller2012}%
  \BibitemOpen
  \bibfield  {author} {\bibinfo {author} {\bibfnamefont {K.}~\bibnamefont
  {Muller}}, \bibinfo {author} {\bibfnamefont {A.}~\bibnamefont {Bechtold}},
  \bibinfo {author} {\bibfnamefont {C.}~\bibnamefont {Ruppert}}, \bibinfo
  {author} {\bibfnamefont {C.}~\bibnamefont {Hautmann}}, \bibinfo {author}
  {\bibfnamefont {J.~S.}\ \bibnamefont {Wildmann}}, \bibinfo {author}
  {\bibfnamefont {T.}~\bibnamefont {Kaldewey}}, \bibinfo {author}
  {\bibfnamefont {M.}~\bibnamefont {Bichler}}, \bibinfo {author} {\bibfnamefont
  {H.~J.}\ \bibnamefont {Krenner}}, \bibinfo {author} {\bibfnamefont
  {G.}~\bibnamefont {Abstreiter}}, \bibinfo {author} {\bibfnamefont
  {M.}~\bibnamefont {Betz}}, \ and\ \bibinfo {author} {\bibfnamefont {J.~J.}\
  \bibnamefont {Finley}},\ }\href {\doibase 10.1103/PhysRevB.85.241306}
  {\bibfield  {journal} {\bibinfo  {journal} {Phys. Rev. B}\ }\textbf {\bibinfo
  {volume} {85}},\ \bibinfo {pages} {241306} (\bibinfo {year}
  {2012})}\BibitemShut {NoStop}%
\bibitem [{\citenamefont {Vamivakas}\ \emph {et~al.}(2009)\citenamefont
  {Vamivakas}, \citenamefont {Zhao}, \citenamefont {Lu},\ and\ \citenamefont
  {Atatuere}}]{Vamivakas2009}%
  \BibitemOpen
  \bibfield  {author} {\bibinfo {author} {\bibfnamefont {A.~N.}\ \bibnamefont
  {Vamivakas}}, \bibinfo {author} {\bibfnamefont {Y.}~\bibnamefont {Zhao}},
  \bibinfo {author} {\bibfnamefont {C.-Y.}\ \bibnamefont {Lu}}, \ and\ \bibinfo
  {author} {\bibfnamefont {M.}~\bibnamefont {Atatuere}},\ }\href {\doibase
  10.1038/NPHYS1182} {\bibfield  {journal} {\bibinfo  {journal} {Nature
  Physics}\ }\textbf {\bibinfo {volume} {5}},\ \bibinfo {pages} {198} (\bibinfo
  {year} {2009})}\BibitemShut {NoStop}%
\bibitem [{\citenamefont {Vamivakas}\ \emph {et~al.}(2010)\citenamefont
  {Vamivakas}, \citenamefont {Lu}, \citenamefont {Matthiesen}, \citenamefont
  {Zhao}, \citenamefont {Falt}, \citenamefont {Badolato},\ and\ \citenamefont
  {Atatuere}}]{Vamivakas2010}%
  \BibitemOpen
  \bibfield  {author} {\bibinfo {author} {\bibfnamefont {A.~N.}\ \bibnamefont
  {Vamivakas}}, \bibinfo {author} {\bibfnamefont {C.~Y.}\ \bibnamefont {Lu}},
  \bibinfo {author} {\bibfnamefont {C.}~\bibnamefont {Matthiesen}}, \bibinfo
  {author} {\bibfnamefont {Y.}~\bibnamefont {Zhao}}, \bibinfo {author}
  {\bibfnamefont {S.}~\bibnamefont {Falt}}, \bibinfo {author} {\bibfnamefont
  {A.}~\bibnamefont {Badolato}}, \ and\ \bibinfo {author} {\bibfnamefont
  {M.}~\bibnamefont {Atatuere}},\ }\href {\doibase 10.1038/nature09359}
  {\bibfield  {journal} {\bibinfo  {journal} {Nature}\ }\textbf {\bibinfo
  {volume} {467}},\ \bibinfo {pages} {297} (\bibinfo {year}
  {2010})}\BibitemShut {NoStop}%
\bibitem [{\citenamefont {Jovanov}\ \emph {et~al.}(2011)\citenamefont
  {Jovanov}, \citenamefont {Kapfinger}, \citenamefont {Bichler}, \citenamefont
  {Abstreiter},\ and\ \citenamefont {Finley}}]{Jovanov2011}%
  \BibitemOpen
  \bibfield  {author} {\bibinfo {author} {\bibfnamefont {V.}~\bibnamefont
  {Jovanov}}, \bibinfo {author} {\bibfnamefont {S.}~\bibnamefont {Kapfinger}},
  \bibinfo {author} {\bibfnamefont {M.}~\bibnamefont {Bichler}}, \bibinfo
  {author} {\bibfnamefont {G.}~\bibnamefont {Abstreiter}}, \ and\ \bibinfo
  {author} {\bibfnamefont {J.~J.}\ \bibnamefont {Finley}},\ }\href {\doibase
  10.1103/PhysRevB.84.235321} {\bibfield  {journal} {\bibinfo  {journal} {Phys.
  Rev. B}\ }\textbf {\bibinfo {volume} {84}},\ \bibinfo {pages} {235321}
  (\bibinfo {year} {2011})}\BibitemShut {NoStop}%
\bibitem [{\citenamefont {Greilich}\ \emph {et~al.}(2009)\citenamefont
  {Greilich}, \citenamefont {Economou}, \citenamefont {Spatzek}, \citenamefont
  {Yakovlev}, \citenamefont {Reuter}, \citenamefont {Wieck}, \citenamefont
  {Reinecke},\ and\ \citenamefont {Bayer}}]{Greilich2009}%
  \BibitemOpen
  \bibfield  {author} {\bibinfo {author} {\bibfnamefont {A.}~\bibnamefont
  {Greilich}}, \bibinfo {author} {\bibfnamefont {S.~E.}\ \bibnamefont
  {Economou}}, \bibinfo {author} {\bibfnamefont {S.}~\bibnamefont {Spatzek}},
  \bibinfo {author} {\bibfnamefont {D.~R.}\ \bibnamefont {Yakovlev}}, \bibinfo
  {author} {\bibfnamefont {D.}~\bibnamefont {Reuter}}, \bibinfo {author}
  {\bibfnamefont {A.~D.}\ \bibnamefont {Wieck}}, \bibinfo {author}
  {\bibfnamefont {T.~L.}\ \bibnamefont {Reinecke}}, \ and\ \bibinfo {author}
  {\bibfnamefont {M.}~\bibnamefont {Bayer}},\ }\href {\doibase
  10.1038/NPHYS1226} {\bibfield  {journal} {\bibinfo  {journal} {Nature
  Physics}\ }\textbf {\bibinfo {volume} {5}},\ \bibinfo {pages} {262} (\bibinfo
  {year} {2009})}\BibitemShut {NoStop}%
\bibitem [{\citenamefont {Press}\ \emph {et~al.}(2008)\citenamefont {Press},
  \citenamefont {Ladd}, \citenamefont {Zhang},\ and\ \citenamefont
  {Yamamoto}}]{Press2008}%
  \BibitemOpen
  \bibfield  {author} {\bibinfo {author} {\bibfnamefont {D.}~\bibnamefont
  {Press}}, \bibinfo {author} {\bibfnamefont {T.~D.}\ \bibnamefont {Ladd}},
  \bibinfo {author} {\bibfnamefont {B.}~\bibnamefont {Zhang}}, \ and\ \bibinfo
  {author} {\bibfnamefont {Y.}~\bibnamefont {Yamamoto}},\ }\href {\doibase
  10.1038/nature07530} {\bibfield  {journal} {\bibinfo  {journal} {Nature}\
  }\textbf {\bibinfo {volume} {456}},\ \bibinfo {pages} {218} (\bibinfo {year}
  {2008})}\BibitemShut {NoStop}%
\bibitem [{\citenamefont {De~Greve}\ \emph {et~al.}(2011)\citenamefont
  {De~Greve}, \citenamefont {McMahon}, \citenamefont {Press}, \citenamefont
  {Ladd}, \citenamefont {Bisping}, \citenamefont {Schneider}, \citenamefont
  {Kamp}, \citenamefont {Worschech}, \citenamefont {Hoefling}, \citenamefont
  {Forchel},\ and\ \citenamefont {Yamamoto}}]{DeGreve2011}%
  \BibitemOpen
  \bibfield  {author} {\bibinfo {author} {\bibfnamefont {K.}~\bibnamefont
  {De~Greve}}, \bibinfo {author} {\bibfnamefont {P.~L.}\ \bibnamefont
  {McMahon}}, \bibinfo {author} {\bibfnamefont {D.}~\bibnamefont {Press}},
  \bibinfo {author} {\bibfnamefont {T.~D.}\ \bibnamefont {Ladd}}, \bibinfo
  {author} {\bibfnamefont {D.}~\bibnamefont {Bisping}}, \bibinfo {author}
  {\bibfnamefont {C.}~\bibnamefont {Schneider}}, \bibinfo {author}
  {\bibfnamefont {M.}~\bibnamefont {Kamp}}, \bibinfo {author} {\bibfnamefont
  {L.}~\bibnamefont {Worschech}}, \bibinfo {author} {\bibfnamefont
  {S.}~\bibnamefont {Hoefling}}, \bibinfo {author} {\bibfnamefont
  {A.}~\bibnamefont {Forchel}}, \ and\ \bibinfo {author} {\bibfnamefont
  {Y.}~\bibnamefont {Yamamoto}},\ }\href {\doibase 10.1038/NPHYS2078}
  {\bibfield  {journal} {\bibinfo  {journal} {Nature Physics}\ }\textbf
  {\bibinfo {volume} {7}},\ \bibinfo {pages} {872} (\bibinfo {year}
  {2011})}\BibitemShut {NoStop}%
\bibitem [{\citenamefont {Xu}\ \emph {et~al.}(2008)\citenamefont {Xu},
  \citenamefont {Sun}, \citenamefont {Berman}, \citenamefont {Steel},
  \citenamefont {Bracker}, \citenamefont {Gammon},\ and\ \citenamefont
  {Sham}}]{Xu2008}%
  \BibitemOpen
  \bibfield  {author} {\bibinfo {author} {\bibfnamefont {X.}~\bibnamefont
  {Xu}}, \bibinfo {author} {\bibfnamefont {B.}~\bibnamefont {Sun}}, \bibinfo
  {author} {\bibfnamefont {P.~R.}\ \bibnamefont {Berman}}, \bibinfo {author}
  {\bibfnamefont {D.~G.}\ \bibnamefont {Steel}}, \bibinfo {author}
  {\bibfnamefont {A.~S.}\ \bibnamefont {Bracker}}, \bibinfo {author}
  {\bibfnamefont {D.}~\bibnamefont {Gammon}}, \ and\ \bibinfo {author}
  {\bibfnamefont {L.~J.}\ \bibnamefont {Sham}},\ }\href {\doibase
  10.1038/nphys1054} {\bibfield  {journal} {\bibinfo  {journal} {Nature
  Physics}\ }\textbf {\bibinfo {volume} {4}},\ \bibinfo {pages} {692} (\bibinfo
  {year} {2008})}\BibitemShut {NoStop}%
\bibitem [{\citenamefont {Kodriano}\ \emph {et~al.}(2012)\citenamefont
  {Kodriano}, \citenamefont {Schwartz}, \citenamefont {Poem}, \citenamefont
  {Benny}, \citenamefont {Presman}, \citenamefont {Truong}, \citenamefont
  {Petroff},\ and\ \citenamefont {Gershoni}}]{Kodriano2012}%
  \BibitemOpen
  \bibfield  {author} {\bibinfo {author} {\bibfnamefont {Y.}~\bibnamefont
  {Kodriano}}, \bibinfo {author} {\bibfnamefont {I.}~\bibnamefont {Schwartz}},
  \bibinfo {author} {\bibfnamefont {E.}~\bibnamefont {Poem}}, \bibinfo {author}
  {\bibfnamefont {Y.}~\bibnamefont {Benny}}, \bibinfo {author} {\bibfnamefont
  {R.}~\bibnamefont {Presman}}, \bibinfo {author} {\bibfnamefont {T.~A.}\
  \bibnamefont {Truong}}, \bibinfo {author} {\bibfnamefont {P.~M.}\
  \bibnamefont {Petroff}}, \ and\ \bibinfo {author} {\bibfnamefont
  {D.}~\bibnamefont {Gershoni}},\ }\href {\doibase 10.1103/PhysRevB.85.241304}
  {\bibfield  {journal} {\bibinfo  {journal} {Phys. Rev. B}\ }\textbf {\bibinfo
  {volume} {85}},\ \bibinfo {pages} {241304} (\bibinfo {year}
  {2012})}\BibitemShut {NoStop}%
\bibitem [{\citenamefont {Borri}\ \emph {et~al.}(2001)\citenamefont {Borri}
  \emph {et~al.}}]{Borri01}%
  \BibitemOpen
  \bibfield  {author} {\bibinfo {author} {\bibfnamefont {P.}~\bibnamefont
  {Borri}} \emph {et~al.},\ }\href {\doibase 10.1103/PhysRevLett.87.157401}
  {\bibfield  {journal} {\bibinfo  {journal} {Phys. Rev. Lett.}\ }\textbf
  {\bibinfo {volume} {87}},\ \bibinfo {pages} {0157401} (\bibinfo {year}
  {2001})}\BibitemShut {NoStop}%
\bibitem [{\citenamefont {Poem}\ \emph {et~al.}(2011)\citenamefont {Poem},
  \citenamefont {Kenneth}, \citenamefont {Kodriano}, \citenamefont {Benny},
  \citenamefont {Khatsevich}, \citenamefont {Avron},\ and\ \citenamefont
  {Gershoni}}]{Poem2011}%
  \BibitemOpen
  \bibfield  {author} {\bibinfo {author} {\bibfnamefont {E.}~\bibnamefont
  {Poem}}, \bibinfo {author} {\bibfnamefont {O.}~\bibnamefont {Kenneth}},
  \bibinfo {author} {\bibfnamefont {Y.}~\bibnamefont {Kodriano}}, \bibinfo
  {author} {\bibfnamefont {Y.}~\bibnamefont {Benny}}, \bibinfo {author}
  {\bibfnamefont {S.}~\bibnamefont {Khatsevich}}, \bibinfo {author}
  {\bibfnamefont {J.~E.}\ \bibnamefont {Avron}}, \ and\ \bibinfo {author}
  {\bibfnamefont {D.}~\bibnamefont {Gershoni}},\ }\href {\doibase
  10.1103/PhysRevLett.107.087401} {\bibfield  {journal} {\bibinfo  {journal}
  {Phys. Rev. Lett.}\ }\textbf {\bibinfo {volume} {107}},\ \bibinfo {pages}
  {087401} (\bibinfo {year} {2011})}\BibitemShut {NoStop}%
\bibitem [{\citenamefont {Kim}\ \emph {et~al.}(2008)\citenamefont {Kim} \emph
  {et~al.}}]{Kim2008}%
  \BibitemOpen
  \bibfield  {author} {\bibinfo {author} {\bibfnamefont {D.}~\bibnamefont
  {Kim}} \emph {et~al.},\ }\href {\doibase 10.1103/PhysRevLett.101.236804}
  {\bibfield  {journal} {\bibinfo  {journal} {Phys. Rev. Lett.}\ }\textbf
  {\bibinfo {volume} {101}},\ \bibinfo {pages} {236804} (\bibinfo {year}
  {2008})}\BibitemShut {NoStop}%
\bibitem [{\citenamefont {Finley}\ \emph {et~al.}(2002)\citenamefont {Finley},
  \citenamefont {Mowbray}, \citenamefont {Skolnick}, \citenamefont {Ashmore},
  \citenamefont {Baker}, \citenamefont {Monte},\ and\ \citenamefont
  {Hopkinson}}]{Finley2002}%
  \BibitemOpen
  \bibfield  {author} {\bibinfo {author} {\bibfnamefont {J.~J.}\ \bibnamefont
  {Finley}}, \bibinfo {author} {\bibfnamefont {D.~J.}\ \bibnamefont {Mowbray}},
  \bibinfo {author} {\bibfnamefont {M.~S.}\ \bibnamefont {Skolnick}}, \bibinfo
  {author} {\bibfnamefont {A.~D.}\ \bibnamefont {Ashmore}}, \bibinfo {author}
  {\bibfnamefont {C.}~\bibnamefont {Baker}}, \bibinfo {author} {\bibfnamefont
  {A.~F.~G.}\ \bibnamefont {Monte}}, \ and\ \bibinfo {author} {\bibfnamefont
  {M.}~\bibnamefont {Hopkinson}},\ }\href {\doibase 10.1103/PhysRevB.66.153316}
  {\bibfield  {journal} {\bibinfo  {journal} {Phys. Rev. B}\ }\textbf {\bibinfo
  {volume} {66}},\ \bibinfo {pages} {153316} (\bibinfo {year}
  {2002})}\BibitemShut {NoStop}%
\bibitem [{\citenamefont {Findeis}\ \emph {et~al.}(2001)\citenamefont
  {Findeis}, \citenamefont {Baier}, \citenamefont {Beham}, \citenamefont
  {Zrenner},\ and\ \citenamefont {Abstreiter}}]{Findeis2001}%
  \BibitemOpen
  \bibfield  {author} {\bibinfo {author} {\bibfnamefont {F.}~\bibnamefont
  {Findeis}}, \bibinfo {author} {\bibfnamefont {M.}~\bibnamefont {Baier}},
  \bibinfo {author} {\bibfnamefont {E.}~\bibnamefont {Beham}}, \bibinfo
  {author} {\bibfnamefont {A.}~\bibnamefont {Zrenner}}, \ and\ \bibinfo
  {author} {\bibfnamefont {G.}~\bibnamefont {Abstreiter}},\ }\href {\doibase
  10.1063/1.1369148} {\bibfield  {journal} {\bibinfo  {journal} {Applied
  Physics Letters}\ }\textbf {\bibinfo {volume} {78}},\ \bibinfo {pages} {2958}
  (\bibinfo {year} {2001})}\BibitemShut {NoStop}%
\bibitem [{\citenamefont {Zrenner}\ \emph {et~al.}(2002)\citenamefont
  {Zrenner}, \citenamefont {Beham}, \citenamefont {Stufler}, \citenamefont
  {Findeis}, \citenamefont {Bichler},\ and\ \citenamefont
  {Abstreiter}}]{Zrenner02}%
  \BibitemOpen
  \bibfield  {author} {\bibinfo {author} {\bibfnamefont {A.}~\bibnamefont
  {Zrenner}}, \bibinfo {author} {\bibfnamefont {E.}~\bibnamefont {Beham}},
  \bibinfo {author} {\bibfnamefont {S.}~\bibnamefont {Stufler}}, \bibinfo
  {author} {\bibfnamefont {F.}~\bibnamefont {Findeis}}, \bibinfo {author}
  {\bibfnamefont {M.}~\bibnamefont {Bichler}}, \ and\ \bibinfo {author}
  {\bibfnamefont {G.}~\bibnamefont {Abstreiter}},\ }\href {\doibase
  10.1038/nature00912} {\bibfield  {journal} {\bibinfo  {journal} {Nature}\
  }\textbf {\bibinfo {volume} {418}},\ \bibinfo {pages} {612} (\bibinfo {year}
  {2002})}\BibitemShut {NoStop}%
\bibitem [{\citenamefont {Fry}\ \emph {et~al.}(2000)\citenamefont {Fry},
  \citenamefont {Finley}, \citenamefont {Wilson}, \citenamefont {Lemaitre},
  \citenamefont {Mowbray}, \citenamefont {Skolnick}, \citenamefont {Hopkinson},
  \citenamefont {Hill},\ and\ \citenamefont {Clark}}]{Fry2000}%
  \BibitemOpen
  \bibfield  {author} {\bibinfo {author} {\bibfnamefont {P.}~\bibnamefont
  {Fry}}, \bibinfo {author} {\bibfnamefont {J.}~\bibnamefont {Finley}},
  \bibinfo {author} {\bibfnamefont {L.}~\bibnamefont {Wilson}}, \bibinfo
  {author} {\bibfnamefont {A.}~\bibnamefont {Lemaitre}}, \bibinfo {author}
  {\bibfnamefont {D.}~\bibnamefont {Mowbray}}, \bibinfo {author} {\bibfnamefont
  {M.}~\bibnamefont {Skolnick}}, \bibinfo {author} {\bibfnamefont
  {M.}~\bibnamefont {Hopkinson}}, \bibinfo {author} {\bibfnamefont
  {G.}~\bibnamefont {Hill}}, \ and\ \bibinfo {author} {\bibfnamefont
  {J.}~\bibnamefont {Clark}},\ }\href {\doibase 10.1063/1.1334363} {\bibfield
  {journal} {\bibinfo  {journal} {Applied Physics Letters}\ }\textbf {\bibinfo
  {volume} {77}},\ \bibinfo {pages} {4344} (\bibinfo {year}
  {2000})}\BibitemShut {NoStop}%
\bibitem [{\citenamefont {M\"uller}\ \emph {et~al.}(2011)\citenamefont
  {M\"uller}, \citenamefont {Reithmaier}, \citenamefont {Clark}, \citenamefont
  {Jovanov}, \citenamefont {Bichler}, \citenamefont {Krenner}, \citenamefont
  {Betz}, \citenamefont {Abstreiter},\ and\ \citenamefont
  {Finley}}]{Mueller2011}%
  \BibitemOpen
  \bibfield  {author} {\bibinfo {author} {\bibfnamefont {K.}~\bibnamefont
  {M\"uller}}, \bibinfo {author} {\bibfnamefont {G.}~\bibnamefont
  {Reithmaier}}, \bibinfo {author} {\bibfnamefont {E.~C.}\ \bibnamefont
  {Clark}}, \bibinfo {author} {\bibfnamefont {V.}~\bibnamefont {Jovanov}},
  \bibinfo {author} {\bibfnamefont {M.}~\bibnamefont {Bichler}}, \bibinfo
  {author} {\bibfnamefont {H.~J.}\ \bibnamefont {Krenner}}, \bibinfo {author}
  {\bibfnamefont {M.}~\bibnamefont {Betz}}, \bibinfo {author} {\bibfnamefont
  {G.}~\bibnamefont {Abstreiter}}, \ and\ \bibinfo {author} {\bibfnamefont
  {J.~J.}\ \bibnamefont {Finley}},\ }\href {\doibase
  10.1103/PhysRevB.84.081302} {\bibfield  {journal} {\bibinfo  {journal} {Phys.
  Rev. B}\ }\textbf {\bibinfo {volume} {84}},\ \bibinfo {pages} {081302}
  (\bibinfo {year} {2011})}\BibitemShut {NoStop}%
\bibitem [{\citenamefont {M\"uller}\ \emph {et~al.}(2012)\citenamefont
  {M\"uller}, \citenamefont {Bechtold}, \citenamefont {Ruppert}, \citenamefont
  {Zecherle}, \citenamefont {Reithmaier}, \citenamefont {Bichler},
  \citenamefont {Krenner}, \citenamefont {Abstreiter}, \citenamefont
  {Holleitner}, \citenamefont {Villas-Boas}, \citenamefont {Betz},\ and\
  \citenamefont {Finley}}]{Mueller2012-1}%
  \BibitemOpen
  \bibfield  {author} {\bibinfo {author} {\bibfnamefont {K.}~\bibnamefont
  {M\"uller}}, \bibinfo {author} {\bibfnamefont {A.}~\bibnamefont {Bechtold}},
  \bibinfo {author} {\bibfnamefont {C.}~\bibnamefont {Ruppert}}, \bibinfo
  {author} {\bibfnamefont {M.}~\bibnamefont {Zecherle}}, \bibinfo {author}
  {\bibfnamefont {G.}~\bibnamefont {Reithmaier}}, \bibinfo {author}
  {\bibfnamefont {M.}~\bibnamefont {Bichler}}, \bibinfo {author} {\bibfnamefont
  {H.~J.}\ \bibnamefont {Krenner}}, \bibinfo {author} {\bibfnamefont
  {G.}~\bibnamefont {Abstreiter}}, \bibinfo {author} {\bibfnamefont {A.~W.}\
  \bibnamefont {Holleitner}}, \bibinfo {author} {\bibfnamefont {J.~M.}\
  \bibnamefont {Villas-Boas}}, \bibinfo {author} {\bibfnamefont
  {M.}~\bibnamefont {Betz}}, \ and\ \bibinfo {author} {\bibfnamefont {J.~J.}\
  \bibnamefont {Finley}},\ }\href {\doibase 10.1103/PhysRevLett.108.197402}
  {\bibfield  {journal} {\bibinfo  {journal} {Phys. Rev. Lett.}\ }\textbf
  {\bibinfo {volume} {108}},\ \bibinfo {pages} {197402} (\bibinfo {year}
  {2012})}\BibitemShut {NoStop}%
\bibitem [{\citenamefont {Zecherle}\ \emph {et~al.}(2010)\citenamefont
  {Zecherle}, \citenamefont {Ruppert}, \citenamefont {Clark}, \citenamefont
  {Abstreiter}, \citenamefont {Finley},\ and\ \citenamefont
  {Betz}}]{Zecherle2010}%
  \BibitemOpen
  \bibfield  {author} {\bibinfo {author} {\bibfnamefont {M.}~\bibnamefont
  {Zecherle}}, \bibinfo {author} {\bibfnamefont {C.}~\bibnamefont {Ruppert}},
  \bibinfo {author} {\bibfnamefont {E.~C.}\ \bibnamefont {Clark}}, \bibinfo
  {author} {\bibfnamefont {G.}~\bibnamefont {Abstreiter}}, \bibinfo {author}
  {\bibfnamefont {J.~J.}\ \bibnamefont {Finley}}, \ and\ \bibinfo {author}
  {\bibfnamefont {M.}~\bibnamefont {Betz}},\ }\href {\doibase
  10.1103/PhysRevB.82.125314} {\bibfield  {journal} {\bibinfo  {journal} {Phys
  Rev B}\ }\textbf {\bibinfo {volume} {82}},\ \bibinfo {pages} {125314}
  (\bibinfo {year} {2010})}\BibitemShut {NoStop}%
\bibitem [{\citenamefont {Benny}\ \emph {et~al.}(2011)\citenamefont {Benny},
  \citenamefont {Khatsevich}, \citenamefont {Kodriano}, \citenamefont {Poem},
  \citenamefont {Presman}, \citenamefont {Galushko}, \citenamefont {Petroff},\
  and\ \citenamefont {Gershoni}}]{Benny2011}%
  \BibitemOpen
  \bibfield  {author} {\bibinfo {author} {\bibfnamefont {Y.}~\bibnamefont
  {Benny}}, \bibinfo {author} {\bibfnamefont {S.}~\bibnamefont {Khatsevich}},
  \bibinfo {author} {\bibfnamefont {Y.}~\bibnamefont {Kodriano}}, \bibinfo
  {author} {\bibfnamefont {E.}~\bibnamefont {Poem}}, \bibinfo {author}
  {\bibfnamefont {R.}~\bibnamefont {Presman}}, \bibinfo {author} {\bibfnamefont
  {D.}~\bibnamefont {Galushko}}, \bibinfo {author} {\bibfnamefont {P.~M.}\
  \bibnamefont {Petroff}}, \ and\ \bibinfo {author} {\bibfnamefont
  {D.}~\bibnamefont {Gershoni}},\ }\href {\doibase
  10.1103/PhysRevLett.106.040504} {\bibfield  {journal} {\bibinfo  {journal}
  {Phys. Rev. Lett.}\ }\textbf {\bibinfo {volume} {106}},\ \bibinfo {pages}
  {040504} (\bibinfo {year} {2011})}\BibitemShut {NoStop}%
\bibitem [{non()}]{non-resonant}%
  \BibitemOpen
  \href@noop {} {}\bibinfo {note} {Non-resonant state control can also be
  achieved by applying the $2\pi$ control pulse to be resonant with the $X-2X$
  transition as in the work of ref. \cite{Poem2011}}\BibitemShut {NoStop}%
\bibitem [{\citenamefont {Economou}\ \emph {et~al.}(2006)\citenamefont
  {Economou}, \citenamefont {Sham}, \citenamefont {Wu},\ and\ \citenamefont
  {Steel}}]{Economou2006}%
  \BibitemOpen
  \bibfield  {author} {\bibinfo {author} {\bibfnamefont {S.~E.}\ \bibnamefont
  {Economou}}, \bibinfo {author} {\bibfnamefont {L.~J.}\ \bibnamefont {Sham}},
  \bibinfo {author} {\bibfnamefont {Y.}~\bibnamefont {Wu}}, \ and\ \bibinfo
  {author} {\bibfnamefont {D.~G.}\ \bibnamefont {Steel}},\ }\href {\doibase
  10.1103/PhysRevB.74.205415} {\bibfield  {journal} {\bibinfo  {journal} {Phys.
  Rev. B}\ }\textbf {\bibinfo {volume} {74}},\ \bibinfo {pages} {205415}
  (\bibinfo {year} {2006})}\BibitemShut {NoStop}%
\bibitem [{\citenamefont {Economou}\ and\ \citenamefont
  {Reinecke}(2007)}]{Economou2007}%
  \BibitemOpen
  \bibfield  {author} {\bibinfo {author} {\bibfnamefont {S.~E.}\ \bibnamefont
  {Economou}}\ and\ \bibinfo {author} {\bibfnamefont {T.~L.}\ \bibnamefont
  {Reinecke}},\ }\href {\doibase 10.1103/PhysRevLett.99.217401} {\bibfield
  {journal} {\bibinfo  {journal} {Phys. Rev. Lett.}\ }\textbf {\bibinfo
  {volume} {99}},\ \bibinfo {pages} {217401} (\bibinfo {year}
  {2007})}\BibitemShut {NoStop}%
\bibitem [{\citenamefont {Takagahara}(2010)}]{Takagahara2010}%
  \BibitemOpen
  \bibfield  {author} {\bibinfo {author} {\bibfnamefont {T.}~\bibnamefont
  {Takagahara}},\ }\href {\doibase 10.1364/JOSAB.27.000A46} {\bibfield
  {journal} {\bibinfo  {journal} {J. Opt. Soc. Am. B}\ }\textbf {\bibinfo
  {volume} {27}},\ \bibinfo {pages} {A46} (\bibinfo {year} {2010})}\BibitemShut
  {NoStop}%
 \bibitem [{\citenamefont {M\"uller}\ \emph {et~al.}(2012)\citenamefont
  {M\"uller}, \citenamefont {Bechtold}, \citenamefont {Ruppert}, \citenamefont
  {Kaldewey}, \citenamefont {Zecherle}, \citenamefont {Wildmann}, \citenamefont
  {Bichler}, \citenamefont {Krener}, \citenamefont {Villas-B\^oas},
  \citenamefont {Abstreiter},\ and\ \citenamefont {Betz}}]{Mueller2012_3}%
  \BibitemOpen
  \bibfield  {author} {\bibinfo {author} {\bibfnamefont {K.}~\bibnamefont
  {M\"uller}}, \bibinfo {author} {\bibfnamefont {A.}~\bibnamefont {Bechtold}},
  \bibinfo {author} {\bibfnamefont {C.}~\bibnamefont {Ruppert}}, \bibinfo
  {author} {\bibfnamefont {T.}~\bibnamefont {Kaldewey}}, \bibinfo {author}
  {\bibfnamefont {M.}~\bibnamefont {Zecherle}}, \bibinfo {author}
  {\bibfnamefont {J.}~\bibnamefont {Wildmann}}, \bibinfo {author}
  {\bibfnamefont {M.}~\bibnamefont {Bichler}}, \bibinfo {author} {\bibfnamefont
  {H.}~\bibnamefont {Krener}}, \bibinfo {author} {\bibfnamefont {J.~M.}\
  \bibnamefont {Villas-B\^oas}}, \bibinfo {author} {\bibfnamefont
  {G.}~\bibnamefont {Abstreiter}}, \ and\ \bibinfo {author} {\bibfnamefont
  {J.~J.}\ \bibnamefont {Betz}, \bibfnamefont {M.and~Finley}},\ }\href
  {\doibase Annalen der Physik, in press} {\  (\bibinfo {year} {2012}),\
  Annalen der Physik, in press}\BibitemShut {NoStop}%
 \bibitem [{\citenamefont {Oulton}\ \emph {et~al.}(2002)\citenamefont {Oulton},
  \citenamefont {Finley}, \citenamefont {Ashmore}, \citenamefont {Gregory},
  \citenamefont {Mowbray}, \citenamefont {Skolnick}, \citenamefont {Steer},
  \citenamefont {Liew}, \citenamefont {Migliorato},\ and\ \citenamefont
  {Cullis}}]{Oulton2002}%
  \BibitemOpen
  \bibfield  {author} {\bibinfo {author} {\bibfnamefont {R.}~\bibnamefont
  {Oulton}}, \bibinfo {author} {\bibfnamefont {J.~J.}\ \bibnamefont {Finley}},
  \bibinfo {author} {\bibfnamefont {A.~D.}\ \bibnamefont {Ashmore}}, \bibinfo
  {author} {\bibfnamefont {I.~S.}\ \bibnamefont {Gregory}}, \bibinfo {author}
  {\bibfnamefont {D.~J.}\ \bibnamefont {Mowbray}}, \bibinfo {author}
  {\bibfnamefont {M.~S.}\ \bibnamefont {Skolnick}}, \bibinfo {author}
  {\bibfnamefont {M.~J.}\ \bibnamefont {Steer}}, \bibinfo {author}
  {\bibfnamefont {S.-L.}\ \bibnamefont {Liew}}, \bibinfo {author}
  {\bibfnamefont {M.~A.}\ \bibnamefont {Migliorato}}, \ and\ \bibinfo {author}
  {\bibfnamefont {A.~J.}\ \bibnamefont {Cullis}},\ }\href {\doibase
  10.1103/PhysRevB.66.045313} {\bibfield  {journal} {\bibinfo  {journal} {Phys.
  Rev. B}\ }\textbf {\bibinfo {volume} {66}},\ \bibinfo {pages} {045313}
  (\bibinfo {year} {2002})}\BibitemShut {NoStop}%
\end{thebibliography}
\end{document}